\documentclass[trackchanges,twocolumn]{aastex701}
\usepackage{amsmath}
\usepackage{graphicx}
\usepackage{caption}
\usepackage{subfigure}

\begin{document}

\title{A Singular Value Decomposition Framework for Jovian Radio Emissions from Parker Solar Probe}

\author[orcid=0009-0008-7739-8359]{Evan Wille}
\affiliation{Astronomy Department, University of California, Berkeley, CA 94720-3411, USA}
\affil{Space Sciences Laboratory, University of California, Berkeley, CA 94720-7450, USA}
\email[show]{ewille@berkeley.edu}  

\author[orcid=0000-0002-0309-9750]{Zack Li}
\affiliation{Berkeley Center for Cosmological Physics, University of California, Berkeley, CA 94720-3411, USA}
\email{zackli@berkeley.edu}

\author[orcid=0000-0002-1573-7457]{Marc Pulupa}
\affil{Space Sciences Laboratory, University of California, Berkeley, CA 94720-7450, USA}
\email{pulupa@berkeley.edu}

\author[orcid=0000-0003-1672-9878]{Philippe Zarka}
\affil{LESIA, Observatoire de Paris, Université PSL, CNRS, 75006 Meudon, France}
\email{philippe.zarka@obspm.fr}

\author[orcid=0009-0003-2581-625X]{L\'{e}on V.E. Koopmans}
\affil{Kapteyn Astronomical Institute, University of Groningen, P.O.Box 800, 9700 AV Groningen, The Netherlands}
\email{koopmans@astro.rug.nl}

\author[0000-0002-1989-3596]{Stuart D. Bale}
\affil{Space Sciences Laboratory, University of California, Berkeley, CA 94720-7450, USA}
\affil{Physics Department, University of California, Berkeley, CA 94720-7300, USA}
\affil{The Blackett Laboratory, Imperial College London, London, SW7 2AZ, UK}
\email{bale@berkeley.edu}

\begin{abstract}
Jovian decametric and hectometric radio emissions provide critical insights into Jupiter’s magnetospheric dynamics and its electrodynamic coupling with Io. Across 21 perihelion encounters since its first light in 2018, the Parker Solar Probe mission has repeatedly recorded high-resolution radio and plasma wave data near closest approach, as a result serving as a long-term distant observer of the Jovian system. However, extracting these relatively faint planetary signals from the continuous solar wind background presents a significant analytical challenge, as the data is routinely saturated by quasi-thermal plasma noise, spacecraft instrumental interference, and solar bursts. To overcome these observational barriers, we present a generalized empirical pipeline that utilizes Singular Value Decomposition coupled with a deterministic, dual-stage noise filtering process. By converting dynamic spectra into a periodic, phase-folded reference frame, this model isolates structured Jovian emissions from the stochastic heliospheric background, demonstrating empirical Signal-to-Noise Ratio improvements of $1.54 \pm 0.11$ dB, reaching up to $3.32$ dB. Additionally, a secondary ``Eigenfaces" matrix factorization is applied to identify long-term global morphological trends and serve as a detection method for future Jovian emissions. This automated mathematical framework extracts transient planetary signals without relying on localized spatial constraints, establishing the viability of PSP as a Jovian radio observatory.
\end{abstract}


\section{Introduction}
    \label{sec:intro}

Jovian decametric and hectometric radio emissions are powerful, complex signatures driven primarily by the Electron Cyclotron Maser Instability within Jupiter's inner magnetosphere. For decades, these highly structured, circularly polarized bursts have served as vital diagnostics for understanding Jupiter's auroral dynamics \citep{1979ApJ...230..621W, 1955JGR....60..213B, 1998JGR...10320159Z} and its intricate electrodynamic coupling with the moon Io \citep{1964Natur.203.1008B, 1980JGR....85.1171N}. Since it began taking measurements in 2018, the Parker Solar Probe has been continuously recording high-resolution solar wind plasma and fields data \citep{2016SSRv..204....7F}. While expressly engineered for heliospheric science and solar proximity, PSP's continuous operational timeline and sensitive radio instrumentation have positioned the spacecraft as a distant, non-dedicated observer of the Jovian system.

Utilizing PSP for planetary radio astronomy, however, presents a significant analytical challenge: extracting relatively faint planetary bursts from a dataset overwhelmingly dominated by solar wind plasma noise. The spacecraft's records are routinely saturated by continuous quasi-thermal noise, severe instrumental interference, and spectrally wide solar bursts. Dedicated planetary observatories and in-situ missions typically handle targeted extractions by relying on close-proximity spatial constraints or localized direction-finding \citep{2005RaSc...40.3003C} to identify target signals. Applying these standard techniques to a distant, non-dedicated platform like PSP is highly inefficient and prone to subjective visual biases, setting up the critical need for a robust, generalized extraction model capable of blindly separating structured planetary signals from a highly dynamic stochastic background.

The objective of this paper is to present an empirical pipeline designed to systematically overcome these observational barriers. By utilizing a mathematical framework based on Singular Value Decomposition coupled with an automated noise filtering process, this model isolates Jovian decametric (DAM) and hectometric (HOM) emissions from the continuous PSP dataset, yielding measurable Signal-to-Noise Ratio improvements across the archival dataset. This approach provides an objective, deterministic method for extracting structured planetary radio bursts from high-noise environments without the need for future manual curation, establishing the viability of PSP as a long-term Jovian radio observatory.

\begin{figure}
    \centering
    \includegraphics[width=1\linewidth]{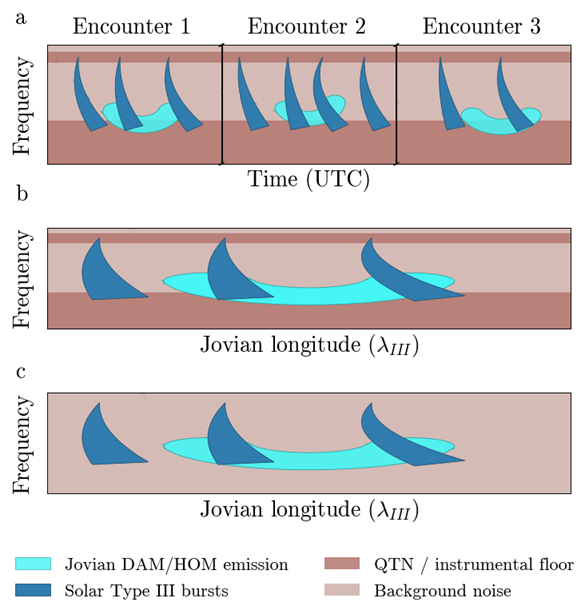}
    \caption{Schematic illustration of the contamination problem addressed by the extraction pipeline and the phase-folding step described in Section 3.1. (a) Raw dynamic spectra across three representative, contiguous PSP encounters, in which transient solar Type II/III bursts (dark blue) and the quasi-thermal-noise/instrumental floor (brown) obscure the underlying Jovian decametric/hectometric emission (light blue). (b) The same encounters re-expressed in the Jovian System III longitude $\lambda_{III}$ reference frame, in which the periodic Jovian signal resolves into a series of coherent, bean-shaped arcs; burst contamination and the QTN/instrumental floor persist unchanged, as phase-folding alone performs no noise removal. (c) The folded spectrum following QTN/instrumental-noise whitening (Section \ref{ssec:data formatting}), isolating the Jovian arcs from the background continuum. Residual burst contamination remains in (c) and is subsequently removed by the SVD framework described in Section \ref{ssec:svd}.}
    \label{fig:placeholder}
\end{figure}

\section{Instrumentation and Data Overview}
    \label{sec:instrumentation}

The FIELDS instrument aboard PSP utilizes a cross-dipole radio antenna consisting of four 2-meter-long electric whip antennas labeled V1-V4 \citep{2016SSRv..204...49B}. Antennas V1 and V2 form the first dipole, while V3 and V4 form the second, rotated by $85^{\circ}$ from the first. Received radio waves are then processed in the Radio Frequency Spectrometer (RFS), which covers the frequency range of 10kHz-19.2MHz and divides received waves into 128 discrete frequency channels. The lowest 64 frequency channel ($10$kHz$-1.7$MHz) are processed in the Low Frequency Receiver (LFR) and highest 64 ($1.3$-$19.2$MHz) are processed in the High Frequency Receiver (HFR). Both LFR and HFR are commandable in cadence: during encounter periods, the $\approx 12$ day intervals when PSP is within $0.25$ AU of the Sun, both receivers produce spectra at a nominal cadence of $\sim 7$s, with a slight time offset between each reciever such that the LFR takes data roughly $3$s after the HFR; outside of encounter, cruise-mode cadence relaxes to $\sim 56$s. Starting in 2023, the cadence increased to once per 3.5s during encounter (up from 7s), and 7s outside of encounter (up from 56s). To maintain a uniform sampling cadence across the dataset, we restrict this analysis to close-encounter-phase intervals. Each $\approx 12$-day encounter already spans dozens of Jupiter's 9.925 hr System III rotations and several of Io's 42.5 hr orbits, providing ample phase coverage; 21 such encounters fall within our October 2018–December 2024 window. The receivers overlap between $1.4-1.7$ MHz so it is important to sort these points to ensure that the frequency channels increase monotonically and resample data to 15s to ensure that data timestamps are equal between LFR and HFR data.

Before it can be used for science, the incoming data is processed via the RFS \citep{2017JGRA..122.2836P}. Antenna signals are processed through a series of analog and digital stages to maintain sensitivity across the intense dynamic range of the solar environment. Received signals from the cross-dipole antennas are first preamplified and managed via a dual-gain strategy: a 50$\times$ high-gain mode for quiet-time signals and a low-gain mode to prevent saturation during intense solar bursts. After analog passband filtering to prevent aliasing, signals are digitized at 38.4 MHz. The onboard FPGA then applies a Polyphase Filter Bank that utilizes a sinc function and Blackman-Harris window to suppress narrowband spacecraft noise by $>100$ dB \citep{2017JGRA..122.2836P}. A 4096-point FFT subsequently processes this data into 2048 frequency bins at full instrumental resolution; because this raw spectral resolution far exceeds what the mission's downlink bandwidth can accommodate \citep{2017JGRA..122.2836P}, these bins are decimated for low-frequency resolution and averaged to reduce statistical noise before compression into 64 pseudo-logarithmically spaced telemetry channels.

PSP performs auto-correlations and calculates the cross-correlation for data taken across the V1V2 and V3V4 pseudo-dipoles. Under ideal conditions, a direction-finding process can extract the Stokes vector $S = [I, Q, U, V]^T$ of a point source by solving the autocorrelation ($A_{XX}$) and cross-correlation ($C_{XY}$) products of the PSP system \citep{2005RaSc...40.3003C, 2014CRPhy..15..441C}:

\begin{equation}
    \label{eq:a}
A_{XX} = \frac{S h_X^2}{2} [ (1+Q)\Omega_X^2 + 2U\Omega_X\Psi_X + (1-Q)\Psi_X^2 ]
\end{equation}

\begin{equation}
\label{eq:cre}
\begin{aligned}
C_{XY}^r = \frac{S h_X h_Y}{2} 
    [&(1+Q)\Omega_X\Omega_Y \\
     &+ U(\Omega_X\Psi_Y + \Omega_Y\Psi_X) \\
     &+ (1-Q)\Psi_X\Psi_Y]
\end{aligned}
\end{equation}

\begin{equation}
\label{eq:cim}
C_{XY}^i = \frac{S h_X h_Y}{2} V [ \Omega_Y\Psi_X - \Omega_X\Psi_Y ]
\end{equation}

where $X$ and $Y$ are the indices for the antennas of interest, $h$ represents the effective antenna length, and $\Omega$ and $\Psi$ are geometric terms dependent on the colatitude and azimuth of the emission source in the sky. An analogous expression for $A_{YY}$ follows by exchanging the X and Y subscripts throughout.

The Jovian extraction methods developed in this paper relies exclusively on $C_{XY}^i$. As shown in Equation \ref{eq:cim}, this component is directly proportional to Stokes $V$. Because Jovian DAM and HOM are highly circularly polarized, this theoretical dependence ideally filters out most unpolarized or linearly polarized interference, isolating the circular Jovian signals as a distinct foreground \citep{1994A&A...286..683D}. Equation \ref{eq:cim} shows that $C^i_{XY}$ is theoretically proportional to Stokes V alone, independent of antenna-source geometry. In practice, however, this exact proportionality assumes the antennas behave as ideal short electrical dipoles, a condition that only holds within a limited frequency range for physical monopole antennas of finite length \citep{2005RaSc...40.3003C}; departures from this short-dipole approximation, phase errors between Antennas $X$ and $Y$, and unfavorable source-antenna geometries allow linearly polarized and nonpolarized components to bleed into $C^i_{XY}$ \citep{1995RaSc...30.1699L}. PSP additionally measures the raw Stokes V polarization of the signal, too, which serve as a proxy for the science conducted with cross-correlation measurements. The full cross-correlation data products, including the directly computed Stokes $V$ polarization, are available only during high-rate telemetry intervals near perihelion, approximately two days per month; outside these windows, no Stokes $V$ data is available. This temporal restriction precludes Stokes $V$ from serving as a continuous long-term polarization monitor and motivates the use of $C^i_{XY}$ as the primary circular polarization proxy throughout this pipeline. No equivalent continuous Stokes $V$ product is available across the HOM band from other heliospheric platforms: Juno-Waves does not produce Stokes parameters directly and is contaminated across the HOM bandpass \citep{2025JGRA..13032826F}; ground-based observatories are limited to DAM frequencies above the ionospheric cutoff ($\gtrsim 10$ MHz); and Wind/WAVES and STEREO-A/WAVES lack the frequency resolution and sensitivity required for continuous HOM monitoring at PSP's cadence. The $C^i_{XY}$ pipeline therefore represents a novel, effective method capable of producing a continuous, broadband circularly polarized HOM dataset across the full PSP archive.

\section{Methods}
    \label{sec:methods}
\subsection{Data Formatting and Noise Reduction}
    \label{ssec:data formatting}
The PSP FIELDS data available after preprocessing is first converted from dynamic spectra to a periodic, phase-folded reference frame relative to Jupiter and Io's orientation from PSP. We use ephemerides from NASA SPICE \citep{1996P&SS...44...65A, 2020JOSS....5.2050A} to determine the real-time geometric relationship for Parker Solar Probe, the Jupiter Barycenter, and Io. For each resampled 15s interval data point, the observation time is first corrected for light travel time between Jupiter and PSP. PSP's position for each data point is projected using SPICE ephemerides onto the Jovian System III reference frame, a rotating coordinate frame with the origin at Jupiter's center and principal axes at the prime meridian and Jupiter's spin axis, to derive PSP and Io's Jovian longitude. We use this reference frame to find $\lambda_{III}$, PSP's Jovian longitude, in addition to Io phase $\Phi_{Io}$, the longitudinal difference of PSP and Io in the same reference frame \citep{2017A&A...604A..17M}. Thus the temporal spectra are folded into two periodic reference frames. This phase-folding restructures the temporal data into repeatable, periodic arrays, establishing the fundamental baseline required for subsequent noise reduction and matrix decomposition. Using the data available between October 2018 and December 2024, 938 individual plots in $\Phi_{Io}$ and 3711 in $\lambda_{III}$ were created.

Selecting and categorizing Jovian data was initially done manually to provide an initial training set to develop the framework; the cross-dipole's gain pattern has very broad lobes at low frequencies (primary lobe having a solid angle of $8\pi /3$, two-thirds of the full sky), so Jupiter is detectable from almost any PSP-Sun-Jupiter geometry except near Sun-Jupiter conjunction \citep{2024A&A...689A.308J}. A manual categorization scheme was used to identify which phase plots contained any Jovian signature; the dominant morphological structures that identified a foreground as Jovian were primarily a distinct, continuous polarity (handedness), attenuation lanes \citep{2015JGRA..120.1888I}, sharing the distinct shapes of vertex-early and vertex-late arcs or Jovian HOM, or having a periodic behavior among multiple successive plots.\footnote{Examples of these behaviors are available in the Appendix} We emphasize that this manual categorization was used solely to construct the initial training set for framework development and to establish ground-truth labels for the validation in Section 4.4; the extraction pipeline itself, rank selection via the entropy threshold (Section \ref{ssec:svd}) and the Eigenface projection score (Section \ref{ssec: eigenfaces}), operates automatically on new data without manual input. Importantly, phase-folding here refers to remapping each datapoint's timestamp onto its corresponding $\lambda_{III}$ or $\Phi_{Io}$ coordinate, not to any temporal averaging or summation; each individual $\approx$12-day encounter is folded and inspected as its own set of dynamic spectra in the new phase coordinate; arc and lane morphologies remain visually intact in this representation exactly as in an unfolded time-frequency spectrogram, only with the horizontal axis remapped from time to phase. This was performed in $\lambda_{III}$ which modulates the morphological structure of such emissions. Of the 3711 plots obtained, 656 had measurable Jovian signatures minimally contaminated by the sun.

\begin{figure*}
    \centering
    \includegraphics[width=0.95\linewidth]{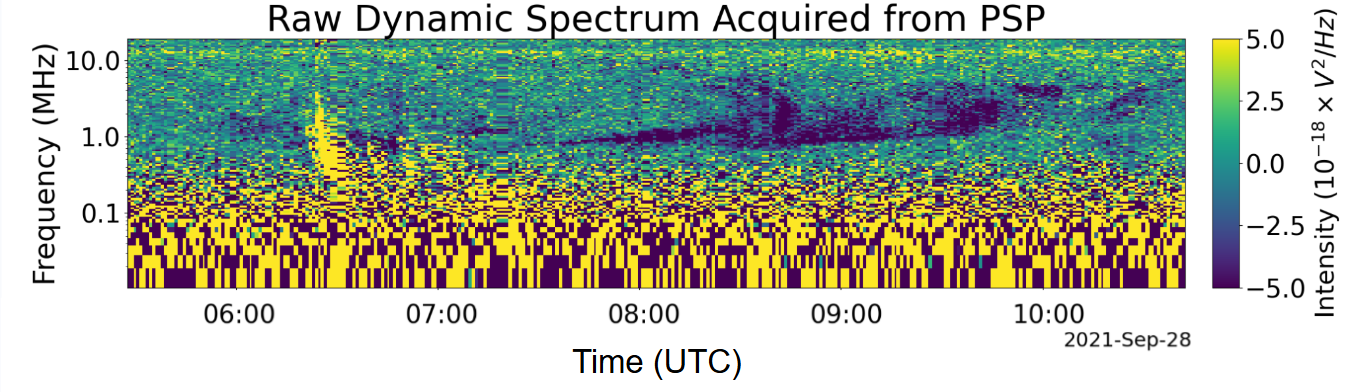}
    \includegraphics[width=0.95\linewidth]{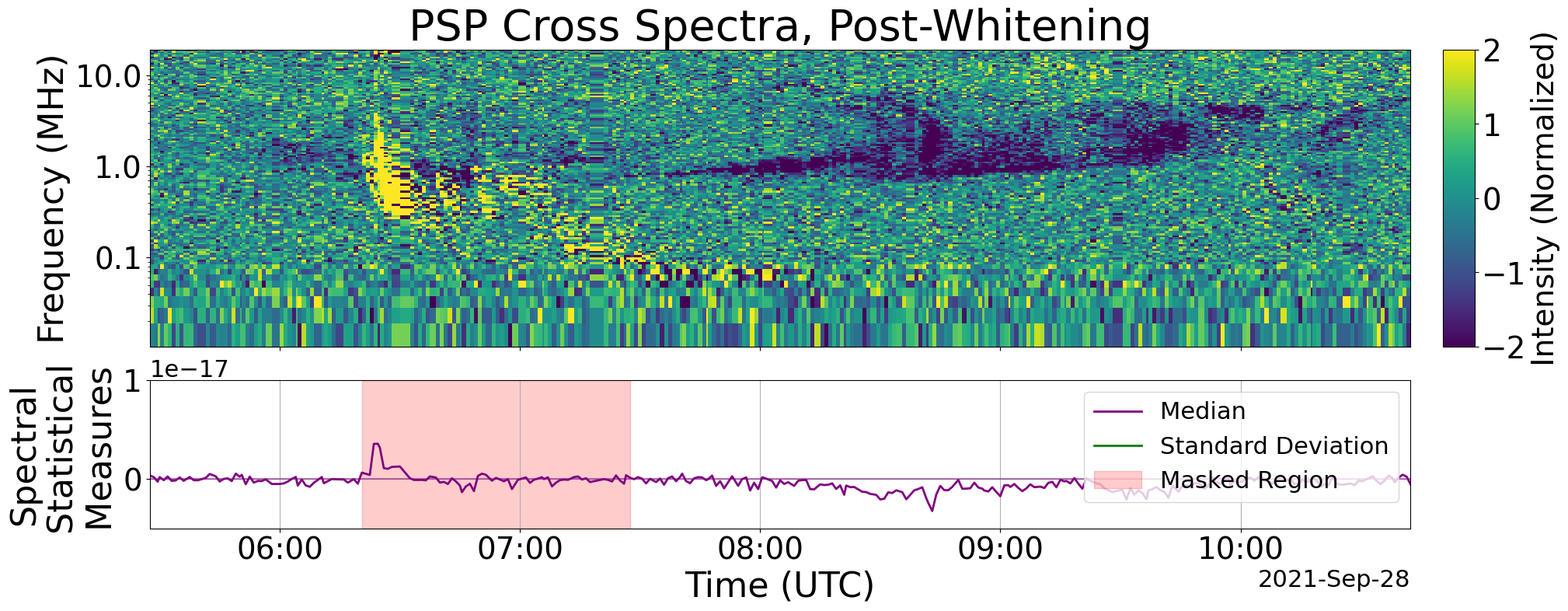}
    \caption{Data formatting and noise reduction applied to a representative PSP cross-spectral interval. Top: The raw dynamic spectrum as acquired from PSP FIELDS/RFS, showing the intense quasi-thermal noise continuum and instrumental interference that dominate the unprocessed data. Bottom: The same interval after dynamic noise-percentile whitening, in which the 75th-percentile frequency-dependent multiplier (Section 3.1) flattens the background noise variance, normalizing the noise level across all channels and revealing embedded structure. The per-column median and standard deviation of the whitened spectrum is shown below, used by the automated thresholding algorithm to detect solar Type II/III burst contamination; the masked region marks an interval flagged by a sustained rise in standard deviation with near-zero mean amplitude, the signature distinguishing broadband solar interference from genuine Jovian emission, which is subsequently removed and replaced via median inpainting.}
    \label{fig:whiten}
\end{figure*}

As seen in the top plot of Figure \ref{fig:whiten}, PSP's dataset is contaminated significantly by local plasma interference at low frequencies, higher-frequency instrumental interference, and transient solar bursts throughout. To isolate the Jovian signal, we start by removing the temporally invariant noise driven by quasi-thermal plasma noise as the frequency channel approaches the local electron plasma frequency in addition to instrumental interference driven by spacecraft-antenna coupling. This baseline continuum is flattened using a dynamic noise-percentile function. By evaluating the median and isolating the 75th percentile of the positive half of the data frequency-by-frequency, we derive a frequency-dependent multiplier to normalize the noise level across all channels. This normalization does not measurably suppress the amplitude of embedded Jovian signal, since the 75th-percentile mask is derived from the noise-dominated majority of each channel and is insensitive to the sparse, transient pixels containing planetary emission. A 75 percentile mask is used here as it is stable across different intensities of quasi-thermal noise without incorporating transient foregrounds in the mask. This dynamic thresholding flattens the background noise along all frequencies to a consistent value and ensures that the localized upper-frequency limit of $f_{pe}$ is masked rather than misclassified as transient Jovian bursts.

In addition to the continuous background noise, Types II and III solar bursts induce severe, spectrally wide transient interference \citep{1985ARA&A..23..169D}. To mitigate this, an automated thresholding technique is employed to detect and flag these spectrally wide transients. At each phase step, the algorithm checks for a sustained increase in the standard deviation across the frequency channels, which indicates a broadband solar burst. Critically, it requires the mean amplitude of the phase column to remain near zero, distinguishing solar interference from genuine Jovian bursts, which inherently increase the mean, as seen in the bottom plot of Figure \ref{fig:whiten} Once identified, these corrupted phase intervals are removed and replaced using a median inpaint scaled to the local background noise floor.

\begin{figure*}
    \centering
    \includegraphics[width=0.85\linewidth]{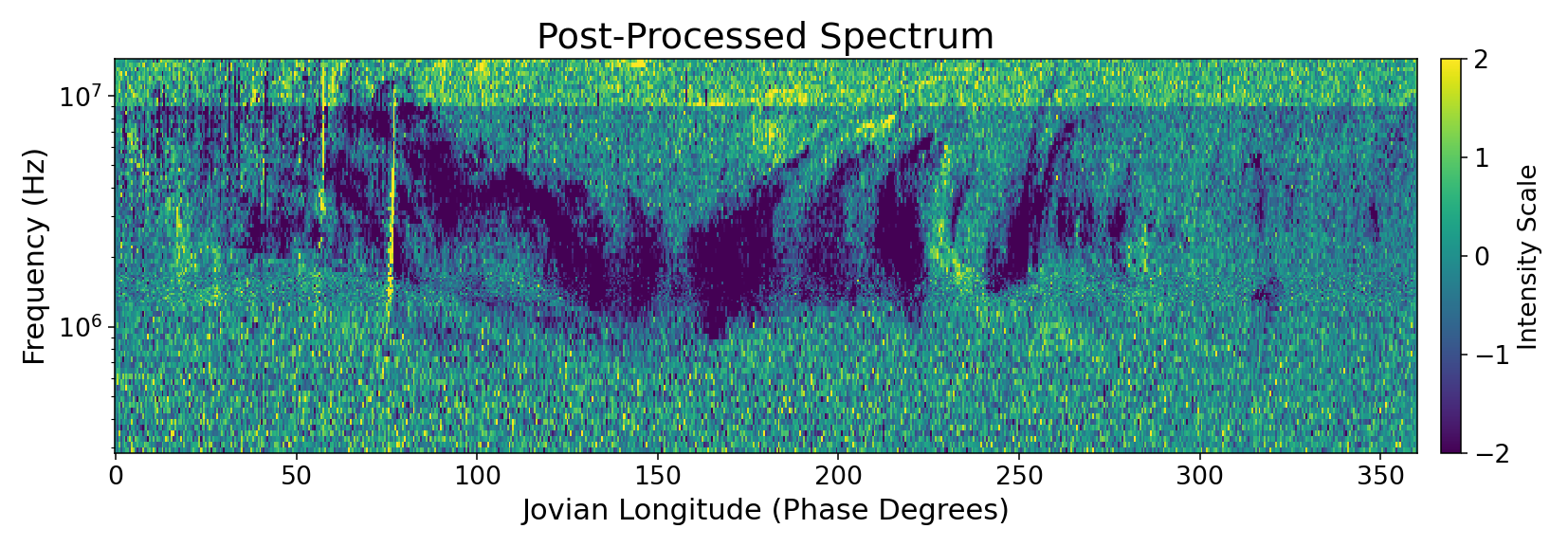}
    \includegraphics[width=0.85\linewidth]{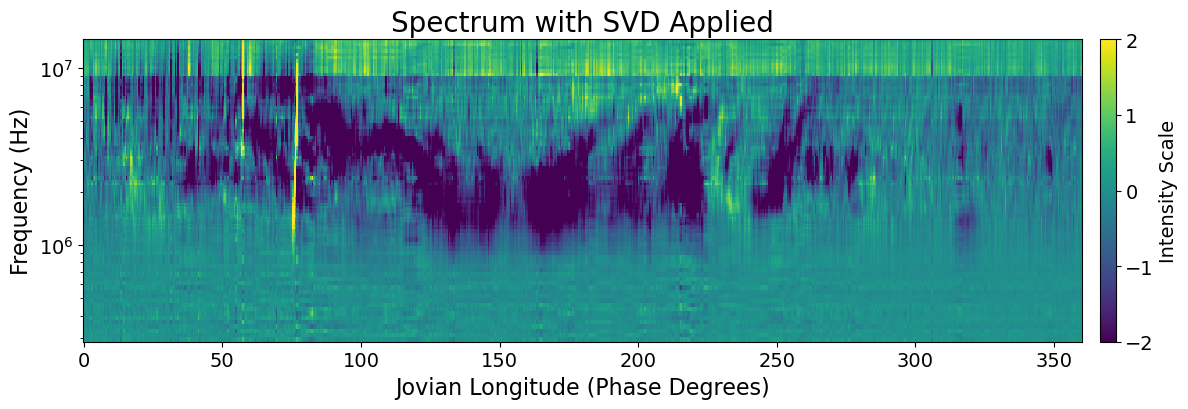}
    \includegraphics[width=0.85\linewidth]{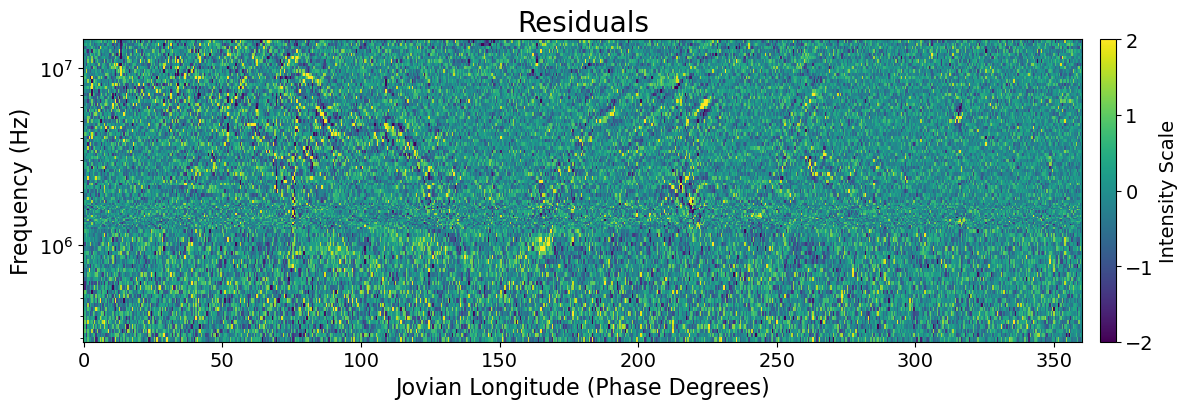}
    \includegraphics[width=0.85\linewidth]{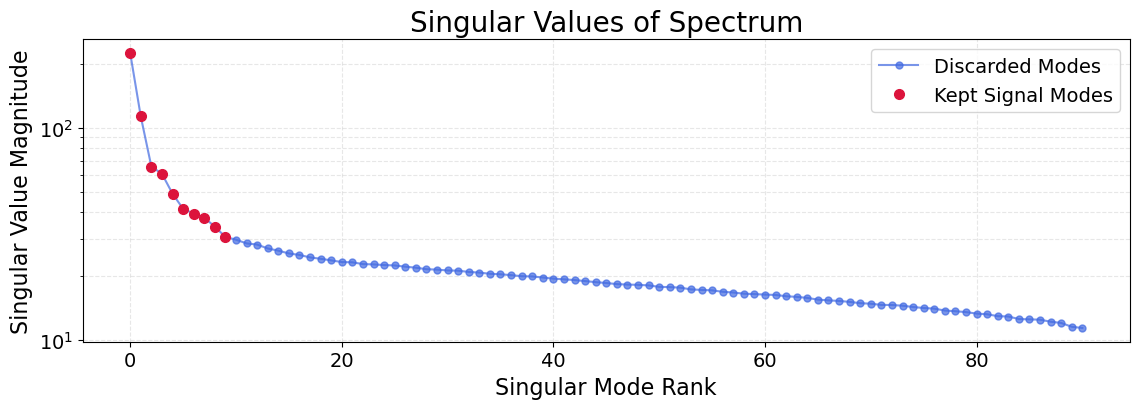}
    \caption{SVD extraction of Jovian decametric and hectometric emissions. \textit{First:} The post-filtered time-frequency spectrogram folded over Jovian System III longitude. \textit{Second:} The reconstructed spectrogram utilizing only the principal signal modes ($k=8$) retained from SVD (all below the preferred entropy limit of $\mathcal{S}=3.0$), demonstrating the successful isolation of structured Jovian emissions (SNR increase of $2.3$ dB for this event). \textit{Third:} The residual matrix ($M_{res}$), consisting primarily of stochastic background noise with slight structured planetary emission remnants. \textit{Fourth:} Distribution of singular values showing the exponential decline in magnitude and the deterministic rank cutoff separating retained signal (red) from discarded noise (blue).} %Additional significant plots of Jovian emissions are available in the Appendix, with the full catalog available through the Github.}
    \label{fig:svd}
\end{figure*}

\subsection{Singular Value Decomposition}
    \label{ssec:svd}
To extract the transient Jovian radio bursts from these post-filtered spectrograms, we apply the matrix factorization method Singular Value Decomposition (SVD). The filtered spectrograms, represented as a matrix $M$, are decomposed into three matrices:

\begin{equation}
\label{eq:svd}
    M=U \Sigma V^T
\end{equation}

where $U$ and $V$ contain the orthogonal spectral and phase basis vectors, respectively, and $\Sigma$ is a diagonal matrix containing the singular values. SVD is utilized because Jovian emissions manifest as highly correlated, structural features across multiple frequency channels and phase bins, whereas the residual background plasma noise is largely uncorrelated \citep{2018ApJ...853..187T, 2023MNRAS.522.1022S}. The structures additionally lack a common spectral and temporal shape, limiting the capability for other extraction methods that need a template to operate well.

According to the Eckart-Young theorem \citep{Eckart1936}, a matrix can be approximated by truncating the decomposition to the first $k$ singular values, such that 

\begin{equation}
\label{eq:eckart young}
    M\approx \sum_{i=1}^kU_i\Sigma_i V^T_i
\end{equation}

The separation of Jovian signals from background noise is evaluated through the distribution of singular values (the diagonal elements of $\Sigma$). In the PSP dataset, this distribution exhibits a steep, quasi-exponential decline, concentrating structured planetary emissions within the principal components before flattening at an inflection point known as the noise limit. To isolate the signal, we establish a deterministic rank parameter $k$ by retaining singular values up to this limit, discarding the remainder.

Following decomposition, each of the singular modes is evaluated using the Shannon entropy \citep{1948BSTJ...27..379S} of its normalized spectral basis vector $\mathbf{u}_i$:

\begin{equation}
\label{eq:shannon}
    \mathcal{S}_i = -\sum_{j} p_{ij} \log p_{ij}, \quad 
    p_{ij} = \frac{|u_{ij}|^2}{\sum_{j}|u_{ij}|^2}.
\end{equation}

Modes dominated by stochastic noise distribute power approximately uniformly across all $N$ frequency channels, producing entropy values approaching the theoretical maximum $\mathcal{S}_{\rm max} = \log N$; for the 90-channel sub-band employed here, $\mathcal{S}_{\rm max} \approx 4.5$ nats (arbitrary, dimensionless units of entropy). Structured Jovian emission modes concentrate power within a subset of channels, yielding substantially lower entropy. A mode is retained in the Eckart-Young reconstruction if and only if $\mathcal{S}_i < \mathcal{S}_{\rm lim}$, where the adopted threshold $\mathcal{S}_{\rm lim} = 3.0$ nats lies well below the noise ceiling. Sensitivity testing and visual inspection across $S_\mathrm{lim} \in [1.0, 4.0]$ nats confirms that the recovered occurrence distributions and emission morphology remain stable over $S_\mathrm{lim} \in [1.5, 3.5]$ with the optimal threshold increasing with the temporal and spectral extent of the structure: larger, more extended emissions require a higher $S_\mathrm{lim}$ to retain the additional modes needed to represent them fully. Degradation seen in visual inspection outside this interval arises from over-rejection at the low end and noise inclusion at the high end, respectively. We note that this absolute threshold is specific to the matrix dimensions used here; application to dynamic spectra of different sizes would require re-calibration of $\mathcal{S}_{\rm lim}$ or normalization relative to $\log N$. While not explored in this project due to the consistent size across the dataset, increasing the size of the dynamic spectra (such as using other instruments with more channels or finer integration times) would increase $\mathcal{S}_{\rm lim}$ under the definitions in \ref{eq:shannon}. The results of this extraction is found in Figure \ref{fig:svd}, with more examples present in the Appendix.

The mathematical efficacy of this SVD extraction is validated through two approaches. First, the separation is quantified by an empirical Signal-to-Noise Ratio (SNR). The empirical SNR at each time-frequency pixel is computed as the ratio of the power at that pixel to a per-channel baseline estimated from the 10th percentile of squared amplitudes across all phase bins in that frequency channel, converted to decibels as $\mathrm{SNR}_{ij} = 10\log_{10}(m_{ij}^2 / b_i)$, where $b_i$ is the baseline for channel $i$. The reported SNR improvement is taken as the difference in mean SNR between the raw and SVD-reconstructed matrices, evaluated only over pixels exceeding a $3$ dB threshold to restrict the comparison to regions containing active emission. Second, the residual matrix $M_{\rm res} = M - M_k$ is retained for quantitative validation in Section \ref{ssec:svd results}, where its spatial correlation structure is analyzed to confirm that the chosen rank $k$ encompasses the Jovian signal without overfitting to the local noise floor.

\begin{figure*}
    \centering
    \includegraphics[width=0.9\linewidth]{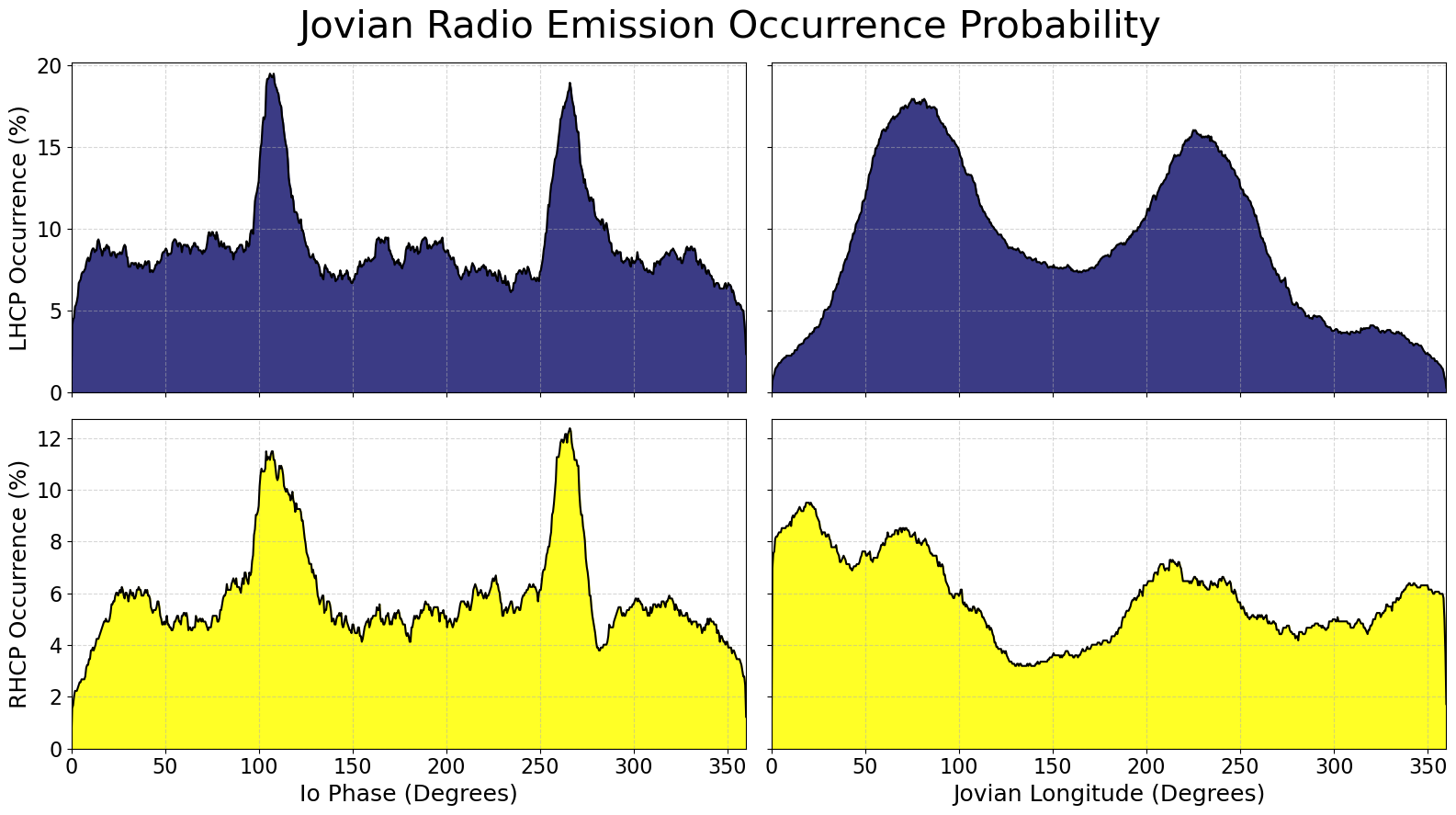}
    \caption{Occurrence probability distributions of SVD-extracted Jovian radio emissions over six years of data in the PSP archive, separated by hemispheric polarization state. Left-hand circularly polarized emissions (LHCP; $C^i_{V1V2-V3V4} < 0$, blue) and right-hand circularly polarized emissions (RHCP; $C^i_{V1V2-V3V4} > 0$, yellow) are shown as a function of Io phase $\Phi_{\rm Io}$ (left column) and Jovian System III longitude $\lambda_{\rm III}$ (right column). The polarization handedness follows the convention used in \cite{1985ARA&A..23..169D}. These polarization states recover concentrated Io-phase peaks at $\Phi_{\rm Io} \approx 100^{\circ}$ and $\approx 260^{\circ}$, consistent with known Io-controlled emission envelopes \citep{2021JGRA..12629435L, 2017A&A...604A..17M}. The longitude distributions exhibit a systematic hemispheric asymmetry, with LHCP showing a smooth bimodal structure and RHCP a flatter, less structured distribution. These features are recovered without any geometric constraint imposed during extraction, providing a geometry-independent validation of the pipeline.}
    \label{fig:occurrence}
\end{figure*}

\subsection{Eigenfaces SVD: Global Trends}
    \label{ssec: eigenfaces}

While the primary SVD pipeline effectively isolates individual, short transients (like in Figure \ref{fig:svd}) from background noise, characterizing recurring morphological trends across the entire dataset requires a secondary matrix factorization approach. For this, we adapt the ``Eigenfaces" methodology, a technique originally developed for pattern identification in facial recognition \citep{1987JOSAA...4..519S, 1991SPIE.1381...43T}. To implement this framework, the extracted individual spectrogram matrices (each with dimensions $N_\nu \times N_\phi$) are flattened into one-dimensional vectors of length $N_\nu N_\phi$. These vectors, drawn from the manually confirmed Jovian-positive subset (Section 3.1), are then stacked to construct a single aggregate matrix used to derive the Eigenface basis. 

A secondary SVD is performed on this aggregate matrix to isolate the dominant global structural modes, or ``Eigenfaces," that represent the fundamental emission topologies present in confirmed Jovian data. Once derived, this basis is applied across the full archive: The prominence of a given structural mode within any specific observation is quantified by a normalized correlation coefficient C, computed as the cosine similarity between the extracted Eigenface vector ($\vec{e}$) and the individual flattened plot vector ($\vec{p}$):
$$C = \frac{\vec{e} \cdot \vec{p}}{|\vec{e}||\vec{p}|}$$
Normalization removes the dependence of the raw dot product on the absolute amplitude of each spectrogram, which otherwise scales with instrumental gain state and burst intensity rather than morphological similarity to the Eigenface template. A negative C indicates a spectrogram anti-correlated with the dominant Eigenface structure; across the archive, anti-correlated intervals are a small minority of both the background (7\%) and Jupiter-confirmed (2\%) populations. We report $|C|$ as the presence metric throughout, consistent with treating anti-correlation as a (rare) alternative expression of structural similarity rather than absence of it. This projection provides a scalar detection metric $C$ for every spectrogram in the archive, quantifying its structural similarity to the dominant emission topology derived from the confirmed training set.

Although this flattening discards the explicit two-dimensional adjacency of the original $N_\nu\times N_\phi$ matrix, the relative ordering of pixels is preserved, so structured time-frequency morphology is retained implicitly in the flattened vector: the Eigenface basis vectors, when reshaped back to $N_\nu \times N_\phi$, recover coherent spatial patterns rather than unstructured noise, consistent with the standard Eigenfaces approach in image recognition \citep{1991SPIE.1381...43T} where 2D images are flattened identically. We note this differs in approach from supervised semantic-segmentation classifiers trained directly on 2D dynamic spectra to identify and localize Jovian decametric emissions on a pixel-by-pixel basis \citep{Aicardi2026}. That method, developed for ground-based NDA observations with a distinct A/B/C/D emission-type classification scheme, reports per-pixel Intersection-over-Union scores of 57-82\% depending on model architecture; our Eigenface metric instead serves as a first-order presence/absence classifier for Jovian emissions in archives the framework is used on rather than a pixel-level classifier. A direct comparison of detection performance between the two approaches when applied to a common dataset is deferred to future work.

The utility and limitations of this metric as an archival detection result are assessed in Section \ref{ssec:eigenface results}.

\section{Results and Discussion}
    \label{sec:results}

\subsection{SVD Extraction Performance}
    \label{ssec:svd results}
The rank parameter $k$ is determined empirically by identifying the inflection point in the singular value spectrum where the steep initial decline transitions to a shallower, approximately linear decay in log-space: the ``noise shoulder'' visible in the bottom plot of Figure \ref{fig:svd}. Singular values above this shoulder concentrate the spatially coherent emission structure, while those below it represent the stochastic noise floor. 50 representative events were manually curated from the confirmed Jovian-positive subset to span the three dominant emission morphologies identified in Section 3.1 (HOM transients, vertex-early arcs, and vertex-late arcs) and to cover a range of emission intensities within each type, rather than drawn via random or stratified sampling. These 50 events are independent of the manual training set used to tune $S_{\rm lim}$, and the reported statistics characterize consistency across representative emission types rather than an archive-wide average. Across these events, the retained rank is $k = 9.56 \pm 0.35$. We emphasize that $k$ is not a fixed operational parameter applied uniformly across the archive; rather, it is determined independently for each phase-folded spectrogram via the entropy-threshold criterion of Section 3.2, retaining all singular modes satisfying $S_i < S_{\rm lim}$. The value reported here is the empirical mean and standard error of this automatically-determined quantity across the 50-event sample, not a value imposed on the pipeline. The low standard error in this measurement indicates that the pipeline consistently isolates Jovian emission structure within a compact, low-dimensional subspace regardless of emission type or intensity. The value of $k$ scales with the spatial extent of the emission in the phase-frequency plane: larger, more extended bursts require more basis vectors to represent their structure fully, while compact or weak emissions are captured in fewer modes.

Across the same 50-event sample, the empirical SNR improves by $1.54 \pm 0.11$ dB on average between the raw and SVD-reconstructed spectrograms, with individual events reaching gains of up to $3.32$ dB. This spread is broadly consistent with the dependence of $k$ on emission extent noted above, as more extended or intense bursts retain additional singular modes and accordingly recover a larger fraction of the structured signal. This behavior is consistent with the SVD truncation acting as an effective denoising operation: because the retained modes are selected via the entropy criterion to concentrate structured signal, the resulting SNR gain is an expected consequence of the method rather than an independent validation of Jovian signal content.

\begin{figure}
    \centering
    \includegraphics[width=1\linewidth]{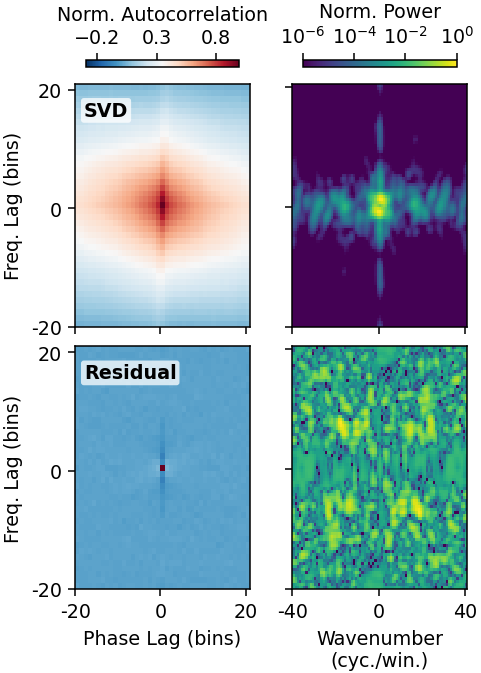}
    \caption{Normalized 2D autocorrelation (left) and corresponding power spectrum (right, Wiener–Khinchin transform of the autocorrelation) of the SVD-reconstructed spectrogram (top) and residual matrix (bottom) for a representative event, evaluated over a ±20 lag-bin window (autocorrelation) and ±40 wavenumber window (power spectrum). The SVD reconstruction exhibits broad spatial correlation consistent with coherent Jovian emission structure concentrated near zero wavenumber in the power spectrum as expected for extended, low-order structure. The residual produces a near-delta-function autocorrelation response; its power spectrum reveals weak, spatially-extended power distributed across a broad range of wavenumbers rather than a single dominant periodicity, indicating that low-level non-white structure remains below the sensitivity of the lag-space autocorrelation alone.}
    \label{fig:autocorrelation}
\end{figure}

To quantitatively validate the separation achieved by the Eckart-Young truncation, we compute the normalized 2D autocorrelation of both the SVD-reconstructed matrix and the residual matrix $M_{\rm res} = M - M_k$ for each event (Figure \ref{fig:autocorrelation}), summarized by the mean normalized autocorrelation in the lag-1 neighborhood of the zero-lag peak. Across the same 50-event sample, the SVD reconstruction yields a lag-1 autocorrelation of $r_{\rm svd} = 0.797 \pm 0.003$, confirming that the retained modes encode genuinely spatially correlated structure consistent with broadband planetary emission. The residual matrix yields $r_{\rm res} = 0.040 \pm 0.017$, close to the theoretical expectation of zero for uncorrelated white noise. The ratio $r_{\rm svd} / r_{\rm res} \approx 20$ quantifies the degree of separation between structured planetary signal and background noise achieved by the pipeline. The 2D autocorrelation of both matrices is additionally transformed via a two-dimensional Hann-windowed Fourier transform to recover the corresponding power spectrum (Wiener–Khinchin theorem), following standard practice in 21-cm intensity-mapping analyses of residual foreground structure. The zero-lag bin, which reflects the total variance of the finite time–frequency window rather than genuine correlation, is set to zero prior to transforming; the Hann taper suppresses the resulting spectral leakage along the wavenumber axes without measurably attenuating off-axis periodic power (verified against injected synthetic test signals).

The near-zero residual autocorrelation additionally confirms that the pipeline does not overfit: if structured emission were being subtracted imperfectly or if noise were being incorporated into the reconstruction, the residual would retain detectable spatial correlation. The consistency of $r_{\rm res}$ across the archive ($\sigma = 0.017$) further indicates that the separation is stable across varying emission intensities and PSP-Jupiter geometries.  occasionally appears along the sharp boundaries of high-intensity emissions \citep{Hansen1987}, a known artifact of low-rank truncation at high-contrast edges; however, the low mean residual autocorrelation confirms these artifacts do not contribute structured power to the background.

Localized ringing (appearance of parallel filaments in the residuals such as $(\lambda_{III}=120^\circ,\nu=4\times10^6\text{ Hz})$ in the Figure \ref{fig:svd} residuals plot) occasionally appears along the sharp boundaries of high-intensity emissions \citep{Hansen1987}, a known artifact of low-rank truncation at high-contrast edges. The power spectrum of the residual autocorrelation (\ref{fig:autocorrelation}, bottom right) confirms this structure is real but weak: power is distributed diffusely across a broad range of spatial wavenumbers rather than concentrated in a single narrowband feature, consistent with a spatially-localized edge artifact rather than a periodic contaminant. This structure contributes negligible power relative to the retained signal (Section \ref{ssec:svd results}), and its absence of a dominant periodicity supports the low mean residual autocorrelation as an adequate summary statistic for archive-wide validation.

\begin{figure*}
    \centering
    \includegraphics[width=0.85\linewidth]{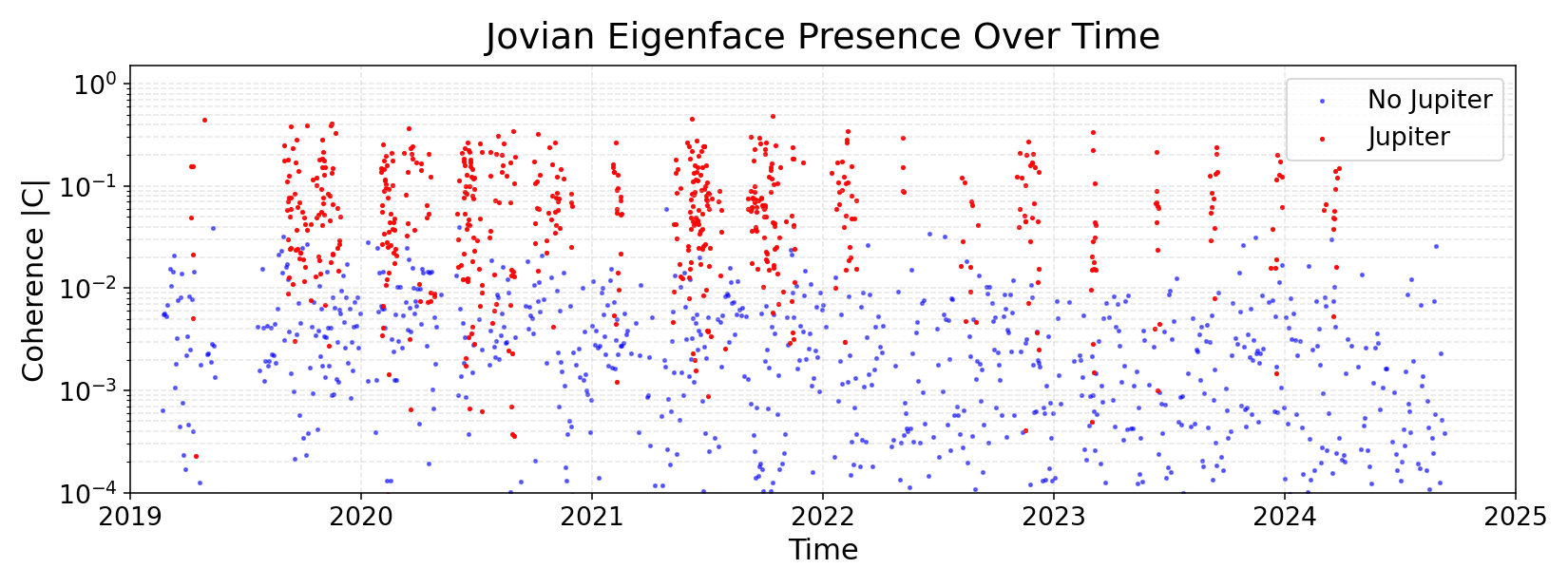}
    \includegraphics[width=0.85\linewidth]{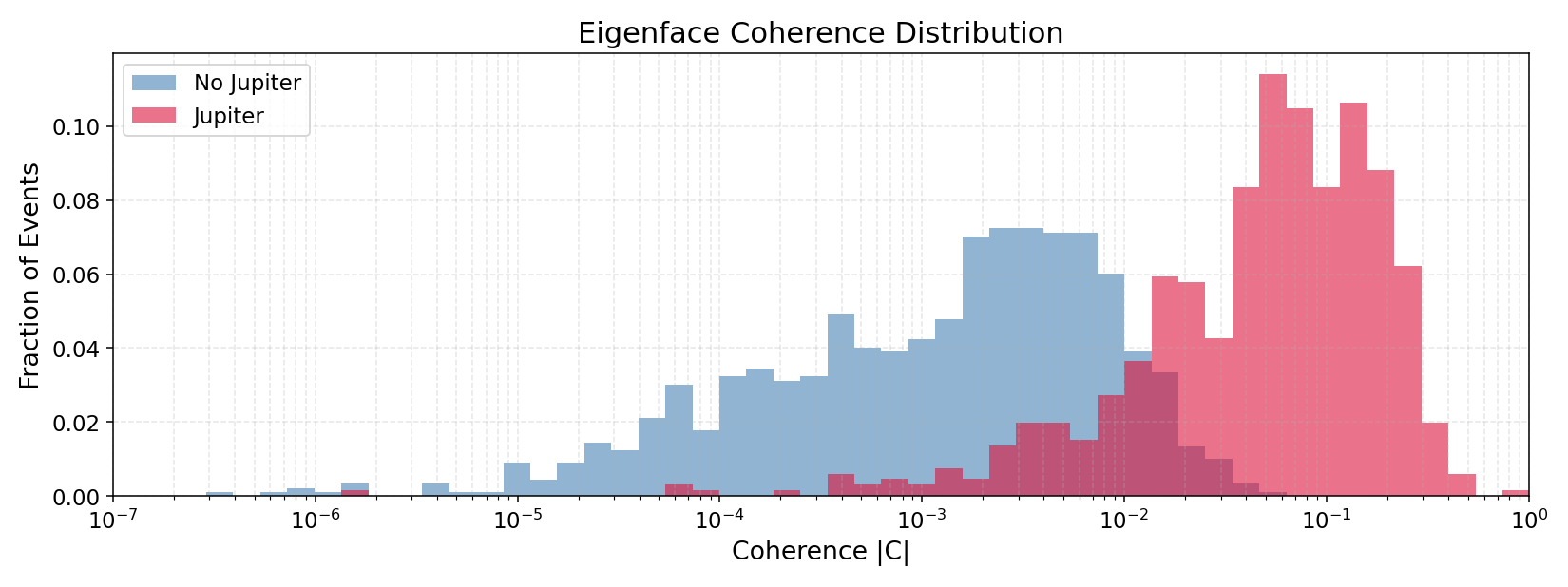}
    \caption{\textit{Top:} Long-term timeline of the Eigenface detection metric $C$ evaluated across a large set of the PSP observational baseline (2019–2025) Emission-confirmed intervals (red) cluster in discrete temporal windows coincident with PSP's perihelion encounters, and are systematically shifted toward higher C than the emission-absent background (blue), though the two populations overlap over an intermediate range of C (Section \ref{ssec:eigenface results}). \textit{Bottom:} Distribution of the normalized Eigenface coherence $|C|$ for emission-absent (``No Jupiter'') and emission-confirmed (``Jupiter'') intervals across the full archive, shown as the fraction of events per log-spaced coherence bin. The two populations are offset in central tendency but overlap substantially over $C \approx 10^{-3}–3\times10^{-2}$, motivating the probabilistic (rather than binary) interpretation of C adopted in Section \ref{ssec:eigenface results}.}
    \label{fig:eigenfaces}
\end{figure*}

\subsection{Validation Against Stokes $V$}
    \label{ssec: stokes v}

To validate that the $C^i_{V1V2-V3V4}$ extraction recovers physically meaningful circularly polarized structure, we compare the SVD-reconstructed $C^i_{V1V2-V3V4}$ spectrogram against an independently processed Stokes $V$ spectrogram over the available high-rate telemetry windows. Both datasets are phase-folded into the same $\lambda_{\rm III}$--$\Phi_{\rm Io}$ reference frame, subject to the same noise-percentile background normalization, and independently processed through the SVD pipeline to suppress stochastic noise prior to comparison. The pixel-wise Pearson correlation coefficient between the two SVD-reconstructed spectrograms, computed across $N = 65{,}611$ phase-frequency pixels, is $r = 0.783$, demonstrating robust morphological agreement between two independent polarization measurements from distinct onboard signal chains.

% \begin{figure}
%     \centering
%     \includegraphics[width=0.95\linewidth]{images/stokes.png}
%     \caption{Pixel-wise comparison of SVD-reconstructed $C^i_{V1V2-V3V4}$ against SVD-reconstructed Stokes $V$ over $N = 65{,}611$ matched  phase-frequency pixels, with color indicating pixel count on a logarithmic scale. The Pearson correlation $r = 0.783$ ($p \ll 10^{-10}$) demonstrates morphological agreement between two independent polarization measurements derived from distinct onboard signal chains. Residual scatter reflects partial bleeding of unpolarized components into both measurements under non-ideal antenna pointing conditions.}
%     \label{fig:stokes}
% \end{figure}

To assess whether the SVD framework's extraction performance derives from the $C^i_{V1V2-V3V4}$ signal chain specifically, or is a generic property of SVD applied to any polarization-sensitive input, we run a control experiment: the identical SVD pipeline is applied directly to the RFS Level 3 Stokes $V$ spectrograms over the same set of available telemetry windows. If SVD extraction performance were independent of input signal chain, both reconstructions should show comparable improvement. Instead, the Stokes $V$ reconstruction yields a lag-1 autocorrelation of $r_{\rm svd,V} = 0.32$ and an SNR improvement of 0.64 dB, both substantially lower than the $r_{\rm svd} = 0.797 \pm 0.003$ and 1.13 dB gain achieved on $C^i_{V1V2-V3V4}$. This gap indicates that the extraction advantage is specific to the input signal chain rather than an artifact of the matrix factorization itself. We attribute this difference to the geometric assumption underlying the RFS Level 3 Stokes $V$ product: its direction-finding deconvolution assumes a source near the solar direction, consistent with the Sun as the RFS' primary calibration target (Section 2). Because Jupiter's position relative to PSP typically differs substantially from the solar direction, this assumption introduces a systematic mismatch that degrades the coherence of the Stokes $V$ spectrogram prior to SVD, leaving less structured signal for the decomposition to recover. The raw $C^i_{V1V2-V3V4}$ product, by contrast, requires no source-location assumption, and remains theoretically proportional to Stokes $V$ (Equation \ref{eq:cim}) for the true Jovian geometry, yielding a cleaner input for low-rank reconstruction. This explanation makes a testable prediction: Stokes V agreement should improve for windows in which Jupiter's angular separation from the Sun, as seen from PSP, is small, since the direction-finding deconvolution's solar-pointing assumption is then closer to correct. We defer a systematic test of this prediction across the archive's high-rate telemetry windows to future work.

\subsection{Occurrence Probability Distributions}
    \label{ssec:occurrence}

To quantitatively assess the fidelity of the SVD extraction pipeline, we compute occurrence probability distributions of the recovered emissions as a function of Jovian System III central meridian longitude ($\lambda_{\rm III}$) and Io phase ($\Phi_{\rm Io}$), the two dominant modulators of Jovian DAM and HOM visibility \citep{1998JGR...10320159Z, 1981Natur.293..382G}. The occurrence probability at each phase bin is the fraction of phase-folded spectrogram intervals flagged as emission-present by the SVD framework, normalized by the total number of valid observations at that bin. Distributions are separated by the sign of $C^i_{V1V2-V3V4}$, isolating left-hand circularly polarized (LHCP; $C^i_{V1V2-V3V4} < 0$) and right-hand circularly polarized (RHCP; $C^i_{V1V2-V3V4} > 0$) emissions into independent occurrence maps, disambiguating overlapping hemispheric source contributions \citep{1994A&A...286..683D, 1998JGR...10320159Z}.

Figure \ref{fig:occurrence} presents each of the four resulting distributions. Both polarization states exhibit two concentrated Io-phase peaks: LHCP peaks near $\Phi_{\rm Io} \approx 100^{\circ}$ and $260^{\circ}$ reaching $\approx 19\%$ occurrence, and RHCP peaks near $\Phi_{\rm Io} \approx 100^{\circ}$ and $265^{\circ}$ reaching $\approx 11\%$ and $12\%$ respectively. Both distributions share a common suppression trough near $\Phi_{\rm Io} \approx 150^{\circ}$, consistent with the known suppression zone documented in prior Jovian emission catalogs \citep{2017A&A...604A..17M, 2021JGRA..12629435L}. The longitude distributions reveal a systematic asymmetry between polarization states: LHCP shows a smooth bimodal structure with maxima near $\lambda_{\rm III} \approx 50^{\circ}$ and $220^{\circ}$, while RHCP exhibits a flatter, less structured distribution. This asymmetry is consistent with hemispherically asymmetric beaming geometries reported in prior statistical studies \citep{1998JGR...10320159Z, 2021JGRA..12629435L}.

The recovery of these phase-space structures --- sharp Io-phase peaks at physically expected longitudes, the inter-peak suppression trough, and the polarization-segregated longitude asymmetry --- without any geometric constraint imposed during extraction constitutes a direct, geometry-independent validation that the pipeline isolates physically coherent Jovian signals rather than noise artifacts. This geometric validation is complemented by the direct polarimetric comparison against independent Stokes $V$ measurements presented in the following section.

\subsection{Eigenface Detection Performance}
    \label{ssec:eigenface results}
To assess the utility of the Eigenfaces decomposition as a large-scale detection tool, we project individual phase-cut dynamic spectra onto the dominant global structural mode derived from the broadband HOM subset of the confirmed training subset (Section \ref{ssec:svd}). The correlation metric $C = \vec{e} \cdot \vec{p}$ provides a scalar presence value for each spectrogram, quantifying its structural similarity to the dominant emission topology encoded in the first Eigenface.

Figure \ref{fig:eigenfaces} presents $C$ evaluated across six years of the PSP observational baseline (top) and the distribution of $C$ for emission-confirmed and emission-absent intervals (bottom). The populations are offset rather than cleanly separated: emission-confirmed intervals peak near $C\approx0.09$ versus $C\approx3\times10^{-3}$ for background ($\sim30\times$ higher), but overlap substantially over $C \approx 10^{-3}-3\times10^{-2}$. $C$ is therefore a probabilistic pre-screen that enriches the candidate pool by $\sim30\times$ rather than a binary classifier; intervals with $C \leq 10^{-4}$ remain reliably emission-absent. Independently, high-$C$ detections cluster into discrete temporal windows coincident with PSP's perihelion encounters (\ref{fig:eigenfaces}), providing a geometry-independent consistency check. This separation is achieved without any prior knowledge of the observation geometry or manual flagging of individual intervals, confirming that the Eigenface projection operates as an effective automated detection screen across the full archive.

The primary utility of this metric is therefore as a computationally efficient first-pass detection flag for large-scale archival screening: intervals returning $C$ above a fixed threshold can be reliably forwarded to the primary SVD pipeline for detailed extraction, while low-$C$ intervals are confidently discarded. We note that the current Eigenface basis does not reliably discriminate between distinct emission morphologies at the available sample sizes, particularly given the limited number of recovered DAM events and residual noise or instrumental contamination. Morphological classification via Eigenface projection is deferred to future work as the PSP archive continues to grow. Similarly, the presence/absence separation achieved by C, while statistically significant and temporally consistent with known encounter geometry, is imperfect at the individual-interval level; the metric is best used as an archival enrichment tool rather than a definitive per-interval classifier.

\begin{figure}
    \centering
    \includegraphics[width=1\linewidth]{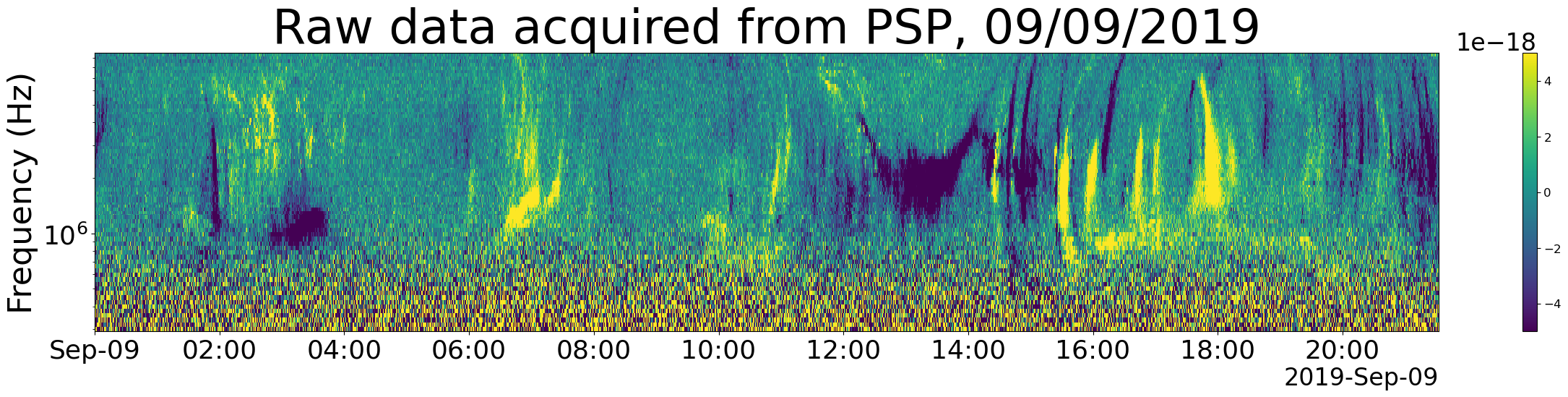}
    \includegraphics[width=1\linewidth]{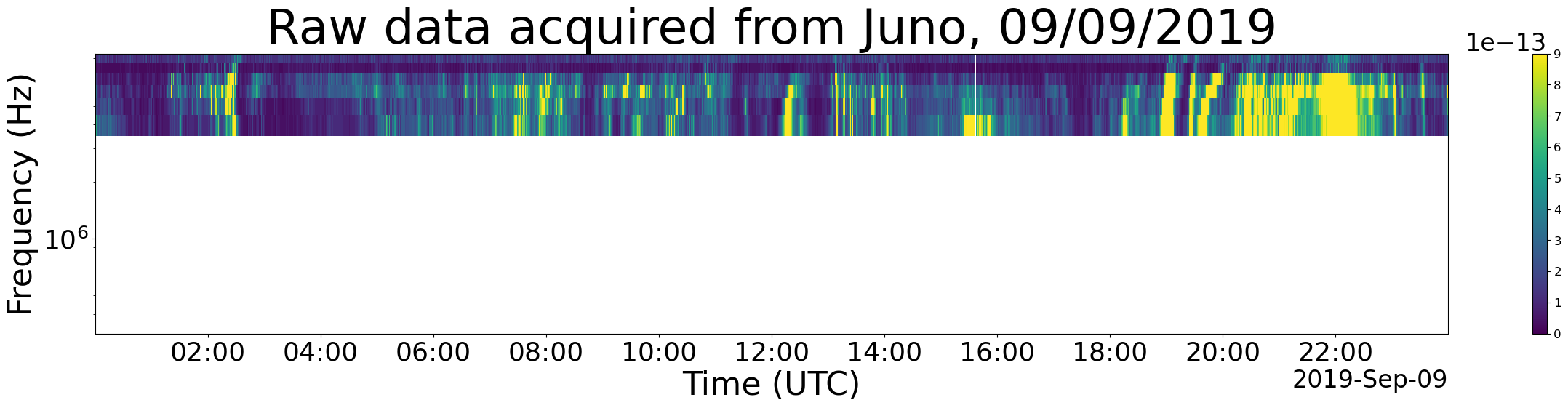}
    \caption{Measurements of the same Jovian HOM transient aboard PSP (top) and Juno (below) at the same frequency band of $300kHz$ to $10MHz$. The Juno data is heavily contaminated by a foreground that prevents the Jovian signatures from being extracted. Nonetheless, the transient is still visible at higher frequencies with a $36.84$min time offset to account for light travel time.}
    \label{fig:juno}
\end{figure}

\section{Complementarity with Juno-Waves}
    \label{ssec:juno}

To provide an independent, geometry-based validation of the PSP-extracted Jovian signal, this section compares simultaneous HOM detections from PSP and the Juno spacecraft, which observes Jupiter's radio emissions in-situ. We show that PSP recovers HOM structure in a frequency band where Juno's own instrumentation is contaminated, and that the timing of coincident detections between the two spacecraft is consistent with the light-travel-time delay expected from their respective distances to Jupiter — a cross-mission check entirely independent of the SVD extraction pipeline itself.

The Juno spacecraft observes Jovian radio emissions in-situ via the Juno-Waves instrument, which measures calibrated electric field spectral density across a nominal bandpass of 50 Hz to 41 MHz \citep{2017SSRv..213..347K, 2015EGUGA..17.6415B}. The survey data product covering the HOM and DAM band (10 kHz-19.2 MHz) is therefore directly comparable in frequency coverage to the PSP RFS. However, a significant portion of Juno's HOM bandpass is rendered unusable by synthesizer electronic interference between 150 kHz and 3.5 MHz \citep{2025JGRA..13032826F} which obscures the majority of hectometric emission that PSP recovers cleanly.

Figure \ref{fig:juno} presents simultaneous measurements of the same HOM transient on 2019 September 9, recorded independently by PSP and Juno. The emission is unrecoverable in Juno's contaminated band but is clearly isolated by the SVD pipeline in the PSP cross-correlation data. Cross-correlating the PSP and Juno lightcurves, collapsed across their shared higher-frequency HOM band, as a function of time lag recovers a peak offset of 38.24 minutes, within 1.4 minutes (3.8\%) of the geometrically predicted light-travel-time delay of $\Delta t = d_{\rm PSP-Jupiter}/c \approx$ 36.84 minutes at the PSP-Jupiter distance of 4.43 AU on that date, providing an independent, quantitative confirmation of the extracted PSP signal's Jovian origin.

PSP therefore complements Juno by providing continuous, broadband HOM coverage in the frequency range most affected by Juno's instrumental contamination. The two datasets are not directly corroborative in a flux-calibration sense, as PSP's cross-correlation measurements and Juno's electric field spectral density require distinct gain correction schemes; a quantitative inter-calibration is deferred to future work.

\section{Conclusion}
    \label{sec:conclusion}
In this study, we presented a generalized empirical methodology to isolate Jovian decametric and hectometric radio emissions from six years of continuous Parker Solar Probe observations. The analytical pipeline separated structured Jovian signals from a stochastic solar wind background through a combination of systematic data formatting, background noise reduction, and primarily the SVD. By converting dynamic spectra into the periodic phase spaces $\lambda_{III}$ and $\Phi_{Io}$ and applying an objective, eigenvalue-based truncation, the model extracted both individual transient Jovian bursts and long-term global morphological trends without relying on localized spatial constraints.

The pipeline's performance is characterized by four quantitative metrics. First, SVD rank selection consistently converges to $k = 9.56 \pm 0.35$ retained modes across 50 manually curated representative events that include the significant modes of emission, demonstrating stable low-dimensional signal representation independent of emission intensity or type. Second, the normalized lag-1 autocorrelation of the residual matrix is $r_{res} = 0.040 \pm 0.017$ across the same sample, consistent with the theoretical expectation for uncorrelated white noise; the corresponding power spectrum (Section \ref{ssec:svd results}) confirms that residual structure, while present at low level, contributes negligible power relative to the retained signal. Third, the SVD-reconstructed $C^i_{V1V2-V3V4}$ achieves a Pearson correlation of $r = 0.783$, $N = 65{,}611$ pixels) against an independent Stokes $V$ reference, establishing cross-instrument polarimetric consistency. Fourth, the polarization-segregated occurrence distributions recover known Io-phase peaks at $\Phi_{\rm Io} \approx 100^{\circ}$ and $260^{\circ}$ without any geometric constraint imposed during extraction, providing an independent geometry-based validation of the pipeline fidelity.

The application of this framework demonstrates that non-dedicated, distant spacecraft can be systematically utilized as viable planetary observatories. Because the SVD extraction relies on the inherent mathematical structure of the emission rather than predefined visual templates or proximity-based direction finding, this open-source methodology is robust. It provides a robust toolset that can be retroactively applied to other heliospheric archival datasets or integrated into upcoming planetary missions to objectively identify and extract faint, transient signals from complex, high-noise environments.

%% Please use the acknowledgment and contribution environments. This will 
%% be anonomyized when the "anonymous" style option is used. 
\begin{acknowledgments}
The PSP/FIELDS experiment was developed and is operated under NASA contract NNN06AA01C.

This project would not have been completed without the additional guidance of Bill Kurth, David Rapetti, Raul Monsalve, and Imke de Pater in providing invaluable scientific insight.

This research used resources of the National Energy Research Scientific Computing Center (NERSC), a Department of Energy User Facility using NERSC award HEP-ERCAP m4895.

S.D.B acknowledges support from the Royal Society Wolfson Visiting Fellowship program.

L.V.E.K. acknowledges the financial support from the European Research Council (ERC) under the European Union's Horizon 2020 research and innovation programme (Grant agreement No. 884760, ``CoDEX").

\end{acknowledgments}

\section*{Data Availability}
The PSP FIELDS/RFS Level 2 and Level 3 data used in this work are publicly available through the FIELDS instrument data archive (\url{https://fields.ssl.berkeley.edu/data/}). Juno-Waves survey data used in Section 5 are available through the NASA Planetary Data System, Planetary Plasma Interactions Node, under Juno WAVES' full-resolution Survey catalog (\url{https://pds-ppi.igpp.ucla.edu/collection/JNO-E_J_SS-WAV-3-CDR-SRVFULL-V2.0:DATA}). SPICE kernels were obtained from the NAIF node of the Planetary Data System. The Python analysis pipeline used to produce the results in this paper is available at \url{https://doi.org/10.5281/zenodo.22075746}. 

\section*{Software}
This work made use of \texttt{NumPy} \citep{Harris2020}, \texttt{SciPy} \citep{Virtanen2020}, \texttt{pandas} \citep{McKinney2010}, and \texttt{Matplotlib} \citep{Hunter2007} for numerical analysis and visualization, and \texttt{Astropy} \citep{Astropy2013, Astropy2018, Astropy2022} for time-system handling. Spacecraft and planetary ephemerides were computed with \texttt{SPICE} \citep{1996P&SS...44...65A} via its Python interface \texttt{SpiceyPy} \citep{2020JOSS....5.2050A}. CDF-formatted PSP and Juno data products were read using \texttt{SpacePy} \citep{Morley2011}. Parallelized processing was performed with \texttt{joblib}, and HDF5 data storage used \texttt{h5py} \citep{Collette2013}. The Python pipeline used to produce the results in this paper is archived on Zenodo \citep{Wille2026psp_jovian_svd}, in addition to the final datasets that the SVD model was performed on. \citep{Wille2026zenodo_dataset}. 

\facility{PSP (FIELDS), Juno (Waves).}
\software{NumPy \citep{Harris2020},
          SciPy \citep{Virtanen2020},
          pandas \citep{McKinney2010},
          Matplotlib \citep{Hunter2007},
          Astropy \citep{Astropy2013,Astropy2018,Astropy2022},
          SpiceyPy \citep{2020JOSS....5.2050A},
          SpacePy \citep{Morley2011},
          h5py \citep{Collette2013},
          joblib,
          psp-jovian-svd \citep{Wille2026psp_jovian_svd}}

\newpage

\bibliographystyle{aasjournalv7}
\bibliography{export-bibtex}

@inproceedings{Aicardi2026,
      author    = {Aicardi, S. and Cecconi, B. and Louis, C. K. and Lamy, L. and Zarka, P.},
      title     = {Deep Learning on {J}ovian Decametric Emissions},
      booktitle = {Planetary, Solar and Heliospheric Radio Emissions X},
      editor    = {Lamy, L. and Louis, C. K. and Fischer, G. and Morosan, D. and Zarka, P.},
      publisher = {OSU Pyth{\'e}as/AMU, Observatoire de Paris},
      year      = {2026},
      doi       = {10.25935/prex-4xv6},
      note      = {Preprint}
}

@ARTICLE{2025JGRA..13032826F,
       author = {{Fischer}, G. and {Taubenschuss}, U. and {P{\'\i}{\v{s}}a}, D. and {Imai}, M. and {Kurth}, W.~S.},
        title = "{Spectral Structures of Jovian Broadband Kilometric Radiation Revealed by Cassini and Juno}",
      journal = {Journal of Geophysical Research (Space Physics)},
         year = 2025,
        month = jan,
       volume = {130},
       number = {1},
        pages = {2024JA032826},
          doi = {10.1029/2024JA032826},
       adsurl = {https://ui.adsabs.harvard.edu/abs/2025JGRA..13032826F}
}

@ARTICLE{2024A&A...689A.308J,
       author = {{J{\'a}come}, H.~R.~P. and {Zarka}, P. and {Louis}, C.~K. and {Marques}, M.~S. and {Echer}, E. and {Lamy}, L.},
        title = "{Effect of the Earth's declination variation on characteristics of Jovian decametric radio emissions}",
      journal = {\aap},
         year = 2024,
        month = sep,
       volume = {689},
          eid = {A308},
        pages = {A308},
          doi = {10.1051/0004-6361/202449868},
       adsurl = {https://ui.adsabs.harvard.edu/abs/2024A&A...689A.308J}
}

@ARTICLE{2023MNRAS.522.1022S,
       author = {{Saxena}, Anchal and {Meerburg}, P. Daniel and {de Lera Acedo}, Eloy and {Handley}, Will and {Koopmans}, L{\'e}on V.~E.},
        title = "{Sky-averaged 21-cm signal extraction using multiple antennas with an SVD framework: the REACH case}",
      journal = {\mnras},
         year = 2023,
        month = jun,
       volume = {522},
       number = {1},
        pages = {1022-1032},
          doi = {10.1093/mnras/stad1047},
archivePrefix = {arXiv},
       eprint = {2212.07415},
 primaryClass = {astro-ph.CO},
       adsurl = {https://ui.adsabs.harvard.edu/abs/2023MNRAS.522.1022S}
}

@ARTICLE{2021JGRA..12629435L,
       author = {{Louis}, C.~K. and {Zarka}, P. and {Dabidin}, K. and {Lampson}, P.-A. and {Magalh{\~a}es}, F.~P. and {Boudouma}, A. and {Marques}, M.~S. and {Cecconi}, B.},
        title = "{Latitudinal Beaming of Jupiter's Radio Emissions From Juno/Waves Flux Density Measurements}",
      journal = {Journal of Geophysical Research (Space Physics)},
         year = 2021,
        month = oct,
       volume = {126},
       number = {10},
          eid = {e29435},
        pages = {e29435},
          doi = {10.1029/2021JA029435},
       adsurl = {https://ui.adsabs.harvard.edu/abs/2021JGRA..12629435L}
}

@ARTICLE{2020JOSS....5.2050A,
       author = {{Annex}, Andrew and {Pearson}, Ben and {Seignovert}, Beno{\^\i}t and {Carcich}, Brian and {Eichhorn}, Helge and {Mapel}, Jesse and {von Forstner}, Johan and {McAuliffe}, Jonathan and {del Rio}, Jorge and {Berry}, Kristin and {Aye}, K.-Michael and {Stefko}, Marcel and {de Val-Borro}, Miguel and {Kulumani}, Shankar and {Murakami}, Shin-ya},
        title = "{SpiceyPy: a Pythonic Wrapper for the SPICE Toolkit}",
      journal = {The Journal of Open Source Software},
         year = 2020,
        month = feb,
       volume = {5},
       number = {46},
          eid = {2050},
        pages = {2050},
          doi = {10.21105/joss.02050},
       adsurl = {https://ui.adsabs.harvard.edu/abs/2020JOSS....5.2050A}
}

@ARTICLE{2018ApJ...853..187T,
       author = {{Tauscher}, Keith and {Rapetti}, David and {Burns}, Jack O. and {Switzer}, Eric},
        title = "{Global 21 cm Signal Extraction from Foreground and Instrumental Effects. I. Pattern Recognition Framework for Separation Using Training Sets}",
      journal = {\apj},
         year = 2018,
        month = feb,
       volume = {853},
       number = {2},
          eid = {187},
        pages = {187},
          doi = {10.3847/1538-4357/aaa41f},
archivePrefix = {arXiv},
       eprint = {1711.03173},
 primaryClass = {astro-ph.IM},
       adsurl = {https://ui.adsabs.harvard.edu/abs/2018ApJ...853..187T}
}

@ARTICLE{2017SSRv..213..347K,
       author = {{Kurth}, W.~S. and {Hospodarsky}, G.~B. and {Kirchner}, D.~L. and {Mokrzycki}, B.~T. and {Averkamp}, T.~F. and {Robison}, W.~T. and {Piker}, C.~W. and {Sampl}, M. and {Zarka}, P.},
        title = "{The Juno Waves Investigation}",
      journal = {\ssr},
         year = 2017,
        month = nov,
       volume = {213},
       number = {1-4},
        pages = {347-392},
          doi = {10.1007/s11214-017-0396-y},
       adsurl = {https://ui.adsabs.harvard.edu/abs/2017SSRv..213..347K}
}

@ARTICLE{2017A&A...604A..17M,
       author = {{Marques}, M.~S. and {Zarka}, P. and {Echer}, E. and {Ryabov}, V.~B. and {Alves}, M.~V. and {Denis}, L. and {Coffre}, A.},
        title = "{Statistical analysis of 26 yr of observations of decametric radio emissions from Jupiter}",
      journal = {\aap},
         year = 2017,
        month = aug,
       volume = {604},
          eid = {A17},
        pages = {A17},
          doi = {10.1051/0004-6361/201630025},
       adsurl = {https://ui.adsabs.harvard.edu/abs/2017A&A...604A..17M}
}

@ARTICLE{2017JGRA..122.2836P,
       author = {{Pulupa}, M. and {Bale}, S.~D. and {Bonnell}, J.~W. and {Bowen}, T.~A. and {Carruth}, N. and {Goetz}, K. and {Gordon}, D. and {Harvey}, P.~R. and {Maksimovic}, M. and {Mart{\'\i}nez-Oliveros}, J.~C. and {Moncuquet}, M. and {Saint-Hilaire}, P. and {Seitz}, D. and {Sundkvist}, D.},
        title = "{The Solar Probe Plus Radio Frequency Spectrometer: Measurement requirements, analog design, and digital signal processing}",
      journal = {Journal of Geophysical Research (Space Physics)},
         year = 2017,
        month = mar,
       volume = {122},
       number = {3},
        pages = {2836-2854},
          doi = {10.1002/2016JA023345},
       adsurl = {https://ui.adsabs.harvard.edu/abs/2017JGRA..122.2836P}
}

@ARTICLE{2016SSRv..204...49B,
       author = {{Bale}, S.~D. and {Goetz}, K. and {Harvey}, P.~R. and {Turin}, P. and {Bonnell}, J.~W. and {Dudok de Wit}, T. and {Ergun}, R.~E. and {MacDowall}, R.~J. and {Pulupa}, M. and {Andre}, M. and {Bolton}, M. and {Bougeret}, J.-L. and {Bowen}, T.~A. and {Burgess}, D. and {Cattell}, C.~A. and {Chandran}, B.~D.~G. and {Chaston}, C.~C. and {Chen}, C.~H.~K. and {Choi}, M.~K. and {Connerney}, J.~E. and {Cranmer}, S. and {Diaz-Aguado}, M. and {Donakowski}, W. and {Drake}, J.~F. and {Farrell}, W.~M. and {Fergeau}, P. and {Fermin}, J. and {Fischer}, J. and {Fox}, N. and {Glaser}, D. and {Goldstein}, M. and {Gordon}, D. and {Hanson}, E. and {Harris}, S.~E. and {Hayes}, L.~M. and {Hinze}, J.~J. and {Hollweg}, J.~V. and {Horbury}, T.~S. and {Howard}, R.~A. and {Hoxie}, V. and {Jannet}, G. and {Karlsson}, M. and {Kasper}, J.~C. and {Kellogg}, P.~J. and {Kien}, M. and {Klimchuk}, J.~A. and {Krasnoselskikh}, V.~V. and {Krucker}, S. and {Lynch}, J.~J. and {Maksimovic}, M. and {Malaspina}, D.~M. and {Marker}, S. and {Martin}, P. and {Martinez-Oliveros}, J. and {McCauley}, J. and {McComas}, D.~J. and {McDonald}, T. and {Meyer-Vernet}, N. and {Moncuquet}, M. and {Monson}, S.~J. and {Mozer}, F.~S. and {Murphy}, S.~D. and {Odom}, J. and {Oliverson}, R. and {Olson}, J. and {Parker}, E.~N. and {Pankow}, D. and {Phan}, T. and {Quataert}, E. and {Quinn}, T. and {Ruplin}, S.~W. and {Salem}, C. and {Seitz}, D. and {Sheppard}, D.~A. and {Siy}, A. and {Stevens}, K. and {Summers}, D. and {Szabo}, A. and {Timofeeva}, M. and {Vaivads}, A. and {Velli}, M. and {Yehle}, A. and {Werthimer}, D. and {Wygant}, J.~R.},
        title = "{The FIELDS Instrument Suite for Solar Probe Plus. Measuring the Coronal Plasma and Magnetic Field, Plasma Waves and Turbulence, and Radio Signatures of Solar Transients}",
      journal = {\ssr},
         year = 2016,
        month = dec,
       volume = {204},
       number = {1-4},
        pages = {49-82},
          doi = {10.1007/s11214-016-0244-5},
       adsurl = {https://ui.adsabs.harvard.edu/abs/2016SSRv..204...49B}
}

@ARTICLE{2016SSRv..204....7F,
       author = {{Fox}, N.~J. and {Velli}, M.~C. and {Bale}, S.~D. and {Decker}, R. and {Driesman}, A. and {Howard}, R.~A. and {Kasper}, J.~C. and {Kinnison}, J. and {Kusterer}, M. and {Lario}, D. and {Lockwood}, M.~K. and {McComas}, D.~J. and {Raouafi}, N.~E. and {Szabo}, A.},
        title = "{The Solar Probe Plus Mission: Humanity's First Visit to Our Star}",
      journal = {\ssr},
         year = 2016,
        month = dec,
       volume = {204},
       number = {1-4},
        pages = {7-48},
          doi = {10.1007/s11214-015-0211-6},
       adsurl = {https://ui.adsabs.harvard.edu/abs/2016SSRv..204....7F}
}

@INPROCEEDINGS{2015EGUGA..17.6415B,
       author = {{Bolton}, Scott and {Owen}, Toby and {Stevenson}, David and {Ingersoll}, Andy and {Connerney}, Jack and {Janssen}, Michael and {Folkner}, William},
        title = "{The Juno Mission and the Origin of Jupiter}",
    booktitle = {EGU General Assembly Conference Abstracts},
         year = 2015,
       series = {EGU General Assembly Conference Abstracts},
        month = apr,
          eid = {6415},
        pages = {6415},
       adsurl = {https://ui.adsabs.harvard.edu/abs/2015EGUGA..17.6415B}
}

@ARTICLE{2015JGRA..120.1888I,
       author = {{Imai}, Masafumi and {Lecacheux}, Alain and {Moncuquet}, Michel and {Bagenal}, Fran and {Higgins}, Charles A. and {Imai}, Kazumasa and {Thieman}, James R.},
        title = "{Modeling Jovian hectometric attenuation lanes during the Cassini flyby of Jupiter}",
      journal = {Journal of Geophysical Research (Space Physics)},
         year = 2015,
        month = mar,
       volume = {120},
       number = {3},
        pages = {1888-1907},
          doi = {10.1002/2014JA020815},
       adsurl = {https://ui.adsabs.harvard.edu/abs/2015JGRA..120.1888I}
}

@ARTICLE{2014CRPhy..15..441C,
       author = {{Cecconi}, Baptiste},
        title = "{Goniopolarimetry: Space-borne radio astronomy with imaging capabilities}",
      journal = {Comptes Rendus Physique},
         year = 2014,
        month = may,
       volume = {15},
       number = {5},
        pages = {441-447},
          doi = {10.1016/j.crhy.2014.02.005},
       adsurl = {https://ui.adsabs.harvard.edu/abs/2014CRPhy..15..441C}
}

@ARTICLE{2005RaSc...40.3003C,
       author = {{Cecconi}, B. and {Zarka}, P.},
        title = "{Direction finding and antenna calibration through analytical inversion of radio measurements performed using a system of two or three electric dipole antennas on a three-axis stabilized spacecraft}",
      journal = {Radio Science},
         year = 2005,
        month = may,
       volume = {40},
       number = {3},
          eid = {RS3003},
        pages = {RS3003},
          doi = {10.1029/2004RS003070},
       adsurl = {https://ui.adsabs.harvard.edu/abs/2005RaSc...40.3003C}
}

@ARTICLE{1998JGR...10320159Z,
       author = {{Zarka}, Philippe},
        title = "{Auroral radio emissions at the outer planets: Observations and theories}",
      journal = {\jgr},
         year = 1998,
        month = sep,
       volume = {103},
       number = {E9},
        pages = {20159-20194},
          doi = {10.1029/98JE01323},
       adsurl = {https://ui.adsabs.harvard.edu/abs/1998JGR...10320159Z}
}

@ARTICLE{1996P&SS...44...65A,
       author = {{Acton}, Charles H.},
        title = "{Ancillary data services of NASA's Navigation and Ancillary Information Facility}",
      journal = {\planss},
         year = 1996,
        month = jan,
       volume = {44},
       number = {1},
        pages = {65-70},
          doi = {10.1016/0032-0633(95)00107-7},
       adsurl = {https://ui.adsabs.harvard.edu/abs/1996P&SS...44...65A}
}

@ARTICLE{1995RaSc...30.1699L,
       author = {{Ladreiter}, H.~P. and {Zarka}, P. and {Lecacheux}, A. and {Macher}, W. and {Rucker}, H.~O. and {Manning}, R. and {Gurnett}, D.~A. and {Kurth}, W.~S.},
        title = "{Analysis of electromagnetic wave direction finding performed by spaceborne antennas using singular-value decomposition techniques}",
      journal = {Radio Science},
         year = 1995,
        month = jan,
       volume = {30},
       number = {6},
        pages = {1699-1712},
          doi = {10.1029/95RS02479},
       adsurl = {https://ui.adsabs.harvard.edu/abs/1995RaSc...30.1699L}
}

@ARTICLE{1994A&A...286..683D,
       author = {{Dulk}, G.~A. and {Leblanc}, Y. and {Lecacheux}, A.},
        title = "{The complete polarization state of Io-related radio storms from Jupiter: a statistical study.}",
      journal = {\aap},
         year = 1994,
        month = jun,
       volume = {286},
        pages = {683-700},
       adsurl = {https://ui.adsabs.harvard.edu/abs/1994A&A...286..683D}
}

@INPROCEEDINGS{1991SPIE.1381...43T,
       author = {{Turk}, Matthew A. and {Pentland}, Alexander P.},
        title = "{Recognition in face space}",
    booktitle = {Intelligent Robots and Computer Vision IX: Algorithms and Techniques},
         year = 1991,
       editor = {{Casasent}, David P.},
       series = {Society of Photo-Optical Instrumentation Engineers (SPIE) Conference Series},
       volume = {1381},
        month = feb,
        pages = {43-54},
          doi = {10.1117/12.25133},
       adsurl = {https://ui.adsabs.harvard.edu/abs/1991SPIE.1381...43T}
}

@ARTICLE{1987JOSAA...4..519S,
       author = {{Sirovich}, L. and {Kirby}, M.},
        title = "{Low-dimensional procedure for the characterization of human faces}",
      journal = {Journal of the Optical Society of America A},
         year = 1987,
        month = mar,
       volume = {4},
       number = {3},
        pages = {519-524},
          doi = {10.1364/JOSAA.4.000519},
       adsurl = {https://ui.adsabs.harvard.edu/abs/1987JOSAA...4..519S}
}

@ARTICLE{1985ARA&A..23..169D,
       author = {{Dulk}, G.~A.},
        title = "{Radio emission from the sun and stars.}",
      journal = {\araa},
         year = 1985,
        month = jan,
       volume = {23},
        pages = {169-224},
          doi = {10.1146/annurev.aa.23.090185.001125},
       adsurl = {https://ui.adsabs.harvard.edu/abs/1985ARA&A..23..169D}
}

@ARTICLE{1981Natur.293..382G,
       author = {{Genova}, F. and {Boischot}, A.},
        title = "{Structure of the source of jovian decametric emission and interplanetary scintillation}",
      journal = {\nat},
         year = 1981,
        month = oct,
       volume = {293},
       number = {5831},
        pages = {382-383},
          doi = {10.1038/293382a0},
       adsurl = {https://ui.adsabs.harvard.edu/abs/1981Natur.293..382G}
}

@ARTICLE{1980JGR....85.1171N,
       author = {{Neubauer}, F.~M.},
        title = "{Nonlinear standing Alfv{\'e}n wave current system at Io: Theory}",
      journal = {\jgr},
         year = 1980,
        month = mar,
       volume = {85},
       number = {A3},
        pages = {1171-1178},
          doi = {10.1029/JA085iA03p01171},
       adsurl = {https://ui.adsabs.harvard.edu/abs/1980JGR....85.1171N}
}

@ARTICLE{1979ApJ...230..621W,
       author = {{Wu}, C.~S. and {Lee}, L.~C.},
        title = "{A theory of the terrestrial kilometric radiation.}",
      journal = {\apj},
         year = 1979,
        month = jun,
       volume = {230},
        pages = {621-626},
          doi = {10.1086/157120},
       adsurl = {https://ui.adsabs.harvard.edu/abs/1979ApJ...230..621W}
}

@ARTICLE{1964Natur.203.1008B,
       author = {{Bigg}, E.~K.},
        title = "{Influence of the Satellite Io on Jupiter's Decametric Emission}",
      journal = {\nat},
         year = 1964,
        month = sep,
       volume = {203},
       number = {4949},
        pages = {1008-1010},
          doi = {10.1038/2031008a0},
       adsurl = {https://ui.adsabs.harvard.edu/abs/1964Natur.203.1008B}
}

@ARTICLE{1955JGR....60..213B,
       author = {{Burke}, B.~F. and {Franklin}, K.~L.},
        title = "{Observations of a Variable Radio Source Associated with the Planet Jupiter}",
      journal = {\jgr},
         year = 1955,
        month = jun,
       volume = {60},
       number = {2},
        pages = {213-217},
          doi = {10.1029/JZ060i002p00213},
       adsurl = {https://ui.adsabs.harvard.edu/abs/1955JGR....60..213B}
}

@ARTICLE{1948BSTJ...27..379S,
       author = {{Shannon}, C.~E.},
        title = "{A mathematical theory of communication}",
      journal = {Bell Labs Technical Journal},
         year = 1948,
        month = jul,
       volume = {27},
       number = {3},
        pages = {379-423},
          doi = {10.1002/j.1538-7305.1948.tb01338.x},
       adsurl = {https://ui.adsabs.harvard.edu/abs/1948BSTJ...27..379S}
}

@article{Eckart1936,
  author  = {Eckart, Carl and Young, Gale},
  title   = {The approximation of one matrix by another of lower rank},
  journal = {Psychometrika},
  year    = {1936},
  volume  = {1},
  number  = {3},
  pages   = {211--218},
  doi     = {10.1007/BF02288367}
}

@article{Hansen1987,
  author  = {Hansen, Per Christian},
  title   = {The truncated {SVD} as a method for regularization},
  journal = {BIT Numerical Mathematics},
  year    = {1987},
  volume  = {27},
  number  = {4},
  pages   = {534--553},
  doi     = {10.1007/BF01937276}
}

@ARTICLE{Harris2020,
       author = {{Harris}, Charles R. and {Millman}, K. Jarrod and {van der Walt}, St{\'e}fan J. and {Gommers}, Ralf and {Virtanen}, Pauli and {Cournapeau}, David and {Wieser}, Eric and {Taylor}, Julian and {Berg}, Sebastian and {Smith}, Nathaniel J. and {Kern}, Robert and {Picus}, Matti and {Hoyer}, Stephan and {van Kerkwijk}, Marten H. and {Brett}, Matthew and {Haldane}, Allan and {del R{\'i}o}, Jaime Fern{\'a}ndez and {Wiebe}, Mark and {Peterson}, Pearu and {G{\'e}rard-Marchant}, Pierre and {Sheppard}, Kevin and {Reddy}, Tyler and {Weckesser}, Warren and {Abbasi}, Hameer and {Gohlke}, Christoph and {Oliphant}, Travis E.},
        title = "{Array programming with NumPy}",
      journal = {Nature},
         year = 2020,
       volume = {585},
        pages = {357-362},
          doi = {10.1038/s41586-020-2649-2}
}

@ARTICLE{Virtanen2020,
       author = {{Virtanen}, Pauli and {Gommers}, Ralf and {Oliphant}, Travis E. and {Haberland}, Matt and {Reddy}, Tyler and {Cournapeau}, David and {Burovski}, Evgeni and {Peterson}, Pearu and {Weckesser}, Warren and {Bright}, Jonathan and {van der Walt}, St{\'e}fan J. and {Brett}, Matthew and {Wilson}, Joshua and {Millman}, K. Jarrod and {Mayorov}, Nikolay and {Nelson}, Andrew R.~J. and {Jones}, Eric and {Kern}, Robert and {Larson}, Eric and {Carey}, C.~J. and {Polat}, {\.I}lhan and {Feng}, Yu and {Moore}, Eric W. and {VanderPlas}, Jake and {Laxalde}, Denis and {Perktold}, Josef and {Cimrman}, Robert and {Henriksen}, Ian and {Quintero}, E.~A. and {Harris}, Charles R. and {Archibald}, Anne M. and {Ribeiro}, Ant{\^o}nio H. and {Pedregosa}, Fabian and {van Mulbregt}, Paul},
        title = "{SciPy 1.0: fundamental algorithms for scientific computing in Python}",
      journal = {Nature Methods},
         year = 2020,
       volume = {17},
        pages = {261-272},
          doi = {10.1038/s41592-019-0686-2}
}

@INPROCEEDINGS{McKinney2010,
       author = {{McKinney}, Wes},
        title = "{Data Structures for Statistical Computing in Python}",
    booktitle = {Proceedings of the 9th Python in Science Conference},
       editor = {{van der Walt}, St{\'e}fan and {Millman}, Jarrod},
         year = 2010,
        pages = {56-61},
          doi = {10.25080/Majora-92bf1922-00a}
}

@ARTICLE{Hunter2007,
       author = {{Hunter}, John D.},
        title = "{Matplotlib: A 2D Graphics Environment}",
      journal = {Computing in Science and Engineering},
         year = 2007,
       volume = {9},
       number = {3},
        pages = {90-95},
          doi = {10.1109/MCSE.2007.55}
}

@ARTICLE{Astropy2013,
       author = {{Astropy Collaboration} and {Robitaille}, T.~P. and {Tollerud}, E.~J. and {Greenfield}, P. and {Droettboom}, M. and {Bray}, E. and {Aldcroft}, T. and {Davis}, M. and {Ginsburg}, A. and {Price-Whelan}, A.~M. and {Kerzendorf}, W.~E. and {Conley}, A. and {Crighton}, N. and {Barbary}, K. and {Muna}, D. and {Ferguson}, H. and {Grollier}, F. and {Parikh}, M.~M. and {Nair}, P.~H. and {Unther}, H.~M. and {Deil}, C. and {Woillez}, J. and {Conseil}, S. and {Kramer}, R. and {Turner}, J.~E.~H. and {Singer}, L. and {Fox}, R. and {Weaver}, B.~A. and {Zabalza}, V. and {Edwards}, Z.~I. and {Azalee Bostroem}, K. and {Burke}, D.~J. and {Casey}, A.~R. and {Crawford}, S.~M. and {Dencheva}, N. and {Ely}, J. and {Jenness}, T. and {Labrie}, K. and {Lim}, P.~L. and {Pierfederici}, F. and {Pontzen}, A. and {Ptak}, A. and {Refsdal}, B. and {Servillat}, M. and {Streicher}, O.},
        title = "{Astropy: A community Python package for astronomy}",
      journal = {Astronomy and Astrophysics},
         year = 2013,
       volume = {558},
          eid = {A33},
          doi = {10.1051/0004-6361/201322068}
}

@ARTICLE{Astropy2018,
       author = {{Astropy Collaboration} and {Price-Whelan}, A.~M. and {Sip{\H{o}}cz}, B.~M. and {G{\"u}nther}, H.~M. and {Lim}, P.~L. and {Crawford}, S.~M. and {Conseil}, S. and {Shupe}, D.~L. and {Craig}, M.~W. and {Dencheva}, N. and {Ginsburg}, A. and {VanderPlas}, J.~T. and {Bradley}, L.~D. and {P{\'e}rez-Su{\'a}rez}, D. and {de Val-Borro}, M. and et al.},
        title = "{The Astropy Project: Building an Open-science Project and Status of the v2.0 Core Package}",
      journal = {The Astronomical Journal},
         year = 2018,
       volume = {156},
          eid = {123},
          doi = {10.3847/1538-3881/aabc4f}
}

@ARTICLE{Astropy2022,
       author = {{Astropy Collaboration} and {Price-Whelan}, A.~M. and {Lim}, P.~L. and {Earl}, N. and {Starkman}, N. and {Bradley}, L. and {Shupe}, D.~L. and {Patil}, A.~A. and {Corrales}, L. and {Brasseur}, C.~E. and {N{\"o}the}, M. and {Donath}, A. and {Tollerud}, E. and {Morris}, B.~M. and {Ginsburg}, A. and et al.},
        title = "{The Astropy Project: Sustaining and Growing a Community-oriented Open-source Project and the Latest Major Release (v5.0) of the Core Package}",
      journal = {The Astrophysical Journal},
         year = 2022,
       volume = {935},
          eid = {167},
          doi = {10.3847/1538-4357/ac7c74}
}

@INPROCEEDINGS{Morley2011,
       author = {{Morley}, S.~K. and {Koller}, J. and {Welling}, D.~T. and {Larsen}, B.~A. and {Henderson}, M.~G. and {Niehof}, J.~T.},
        title = "{SpacePy - A Python-based Library of Tools for the Space Sciences}",
    booktitle = {Proceedings of the 9th Python in Science Conference (SciPy 2010)},
      address = {Austin, TX},
         year = 2011,
        pages = {67-72},
          doi = {10.25080/Majora-92bf1922-012}
}

@BOOK{Collette2013,
       author = {{Collette}, Andrew},
        title = "{Python and HDF5}",
    publisher = {O'Reilly Media},
         year = 2013,
      address = {Sebastopol, CA}
}

@software{Wille2026psp_jovian_svd,
  author       = {Wille, Evan and Li, Zack and Pulupa, Marc and Zarka, Philippe and Koopmans, Leon V. E. and Bale, Stuart D.},
  title        = {{psp-jovian-svd: An SVD Framework for Jovian Radio Emissions from PSP}},
  year         = {2026},
  publisher    = {Zenodo},
  version      = {v1.0-agu-submission},
  doi          = {10.5281/zenodo.22075746},
  url          = {https://doi.org/10.5281/zenodo.22075746}
}

@misc{Wille2026zenodo_dataset,
  author       = {Wille, Evan and Li, Zack and Pulupa, Marc and Zarka, Philippe and Koopmans, Leon and Bale, Stuart},
  title        = {{Supplemental datasets and SPICE kernels used for "An SVD Framework for Jovian Radio Emissions from PSP"}},
  year         = {2026},
  month        = aug,
  publisher    = {Zenodo},
  version      = {v1},
  doi          = {10.5281/zenodo.22076017},
  url          = {https://doi.org/10.5281/zenodo.22076017}
}

\vspace{1cm}

\appendix

\section{Selected Plots}
\begin{figure*}[p]
    \centering
    \includegraphics[width=0.85\linewidth]{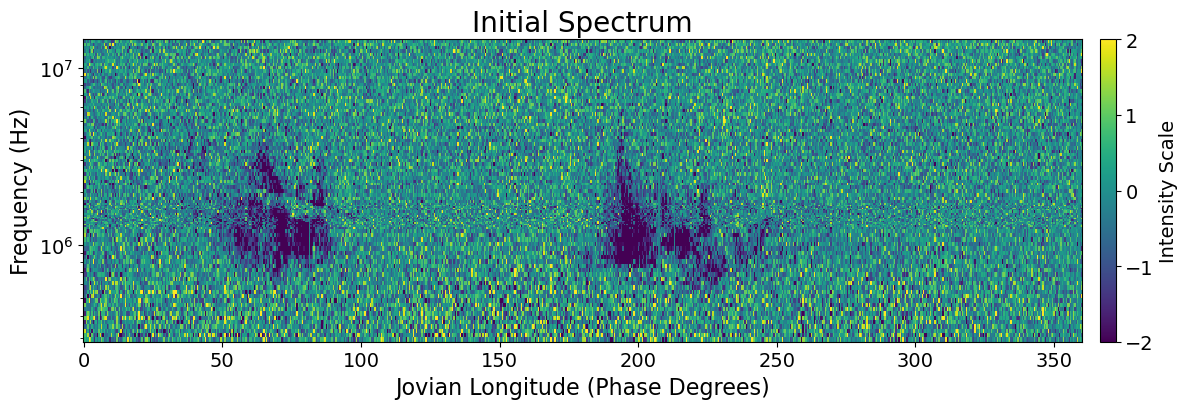}
    \includegraphics[width=0.85\linewidth]{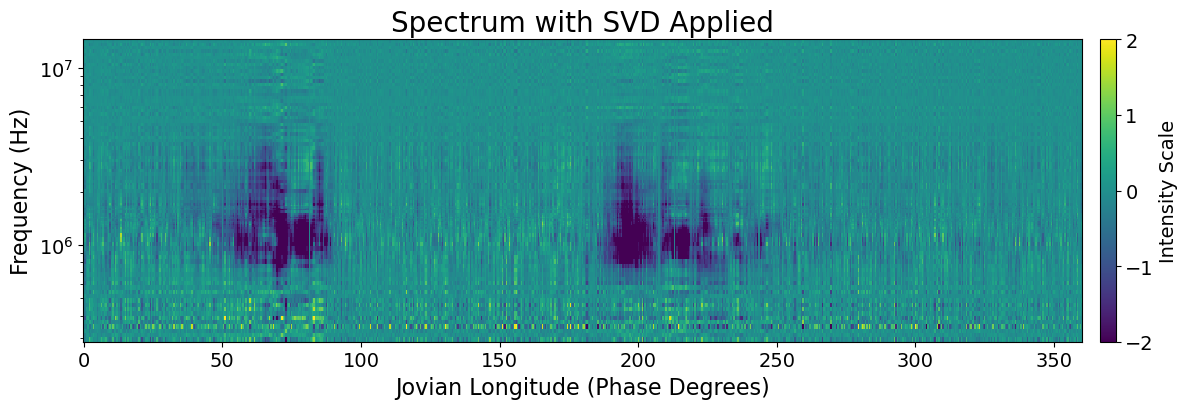}
    \includegraphics[width=0.85\linewidth]{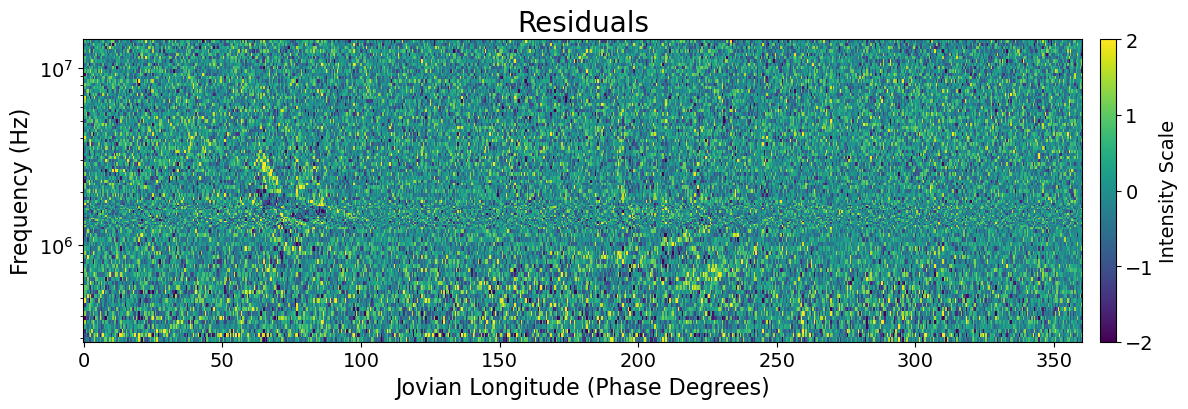}
    \includegraphics[width=0.85\linewidth]{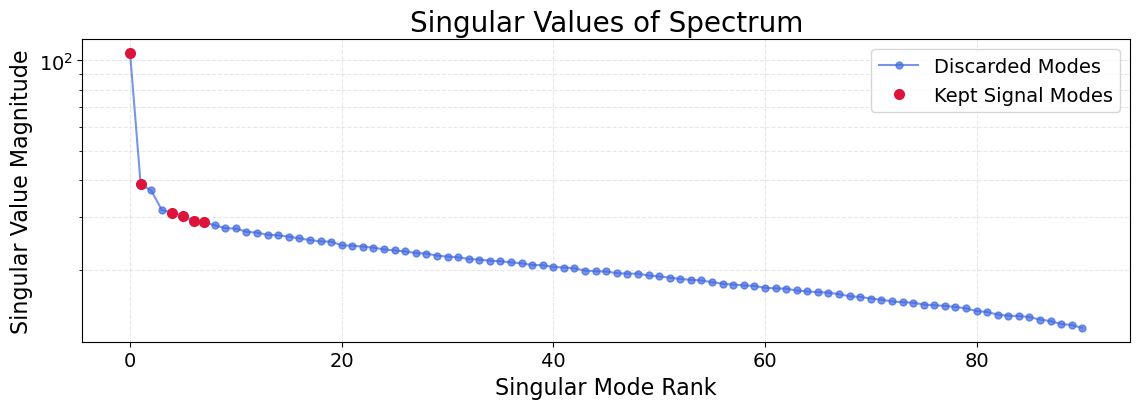}
    \caption{SVD extraction of a quieter pair of Jovian hectometric emissions.}
    \label{app1}
\end{figure*}

\begin{figure*}
    \centering
    \includegraphics[width=0.85\linewidth]{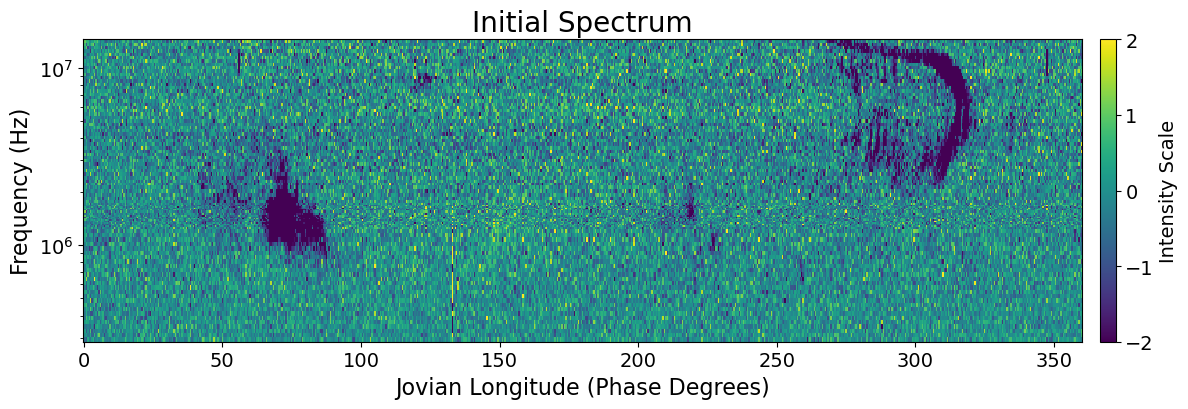}
    \includegraphics[width=0.85\linewidth]{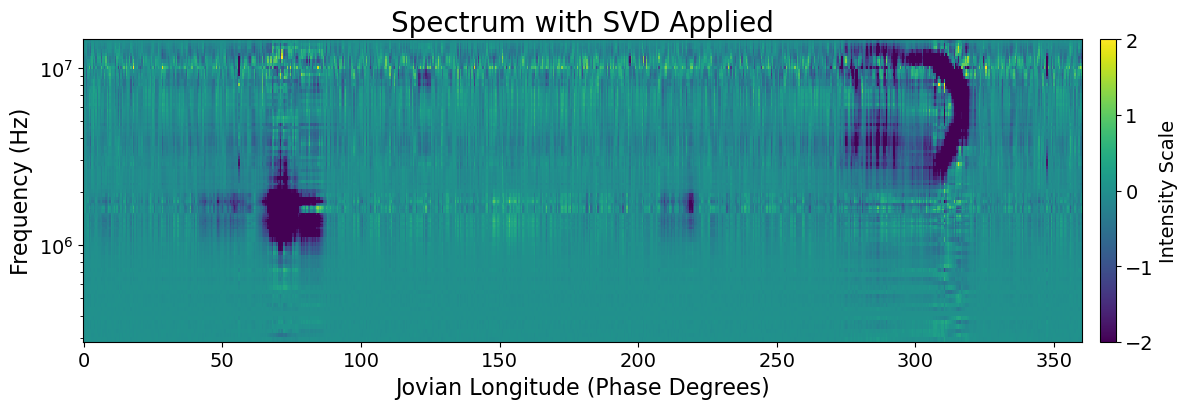}
    \includegraphics[width=0.85\linewidth]{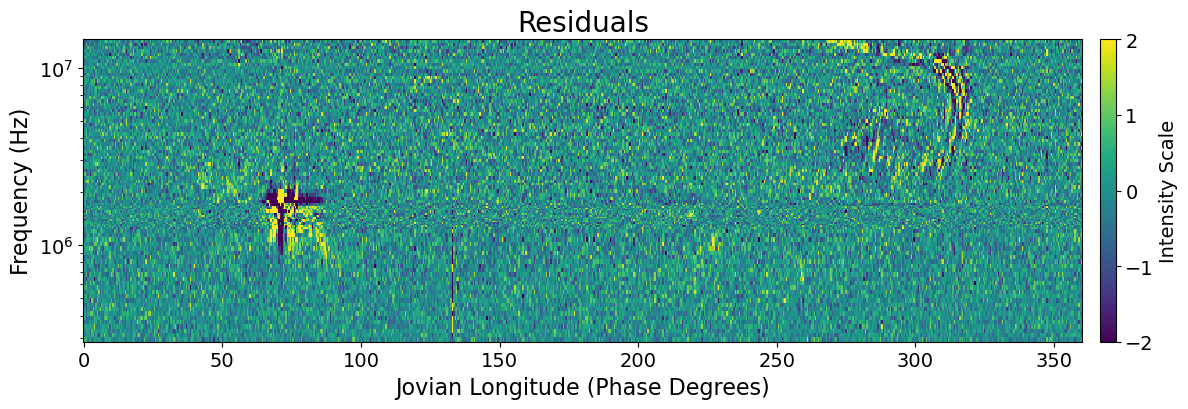}
    \includegraphics[width=0.85\linewidth]{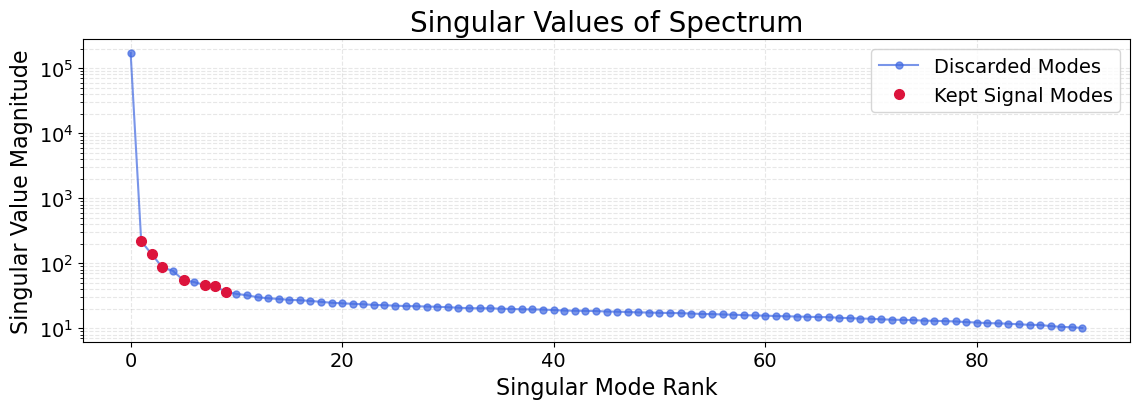}
    \caption{SVD extraction of a decametric vertex-late burst (the arc on the right of the plots).}
    \label{app2}
\end{figure*}

\begin{figure*}
    \centering
    \includegraphics[width=0.85\linewidth]{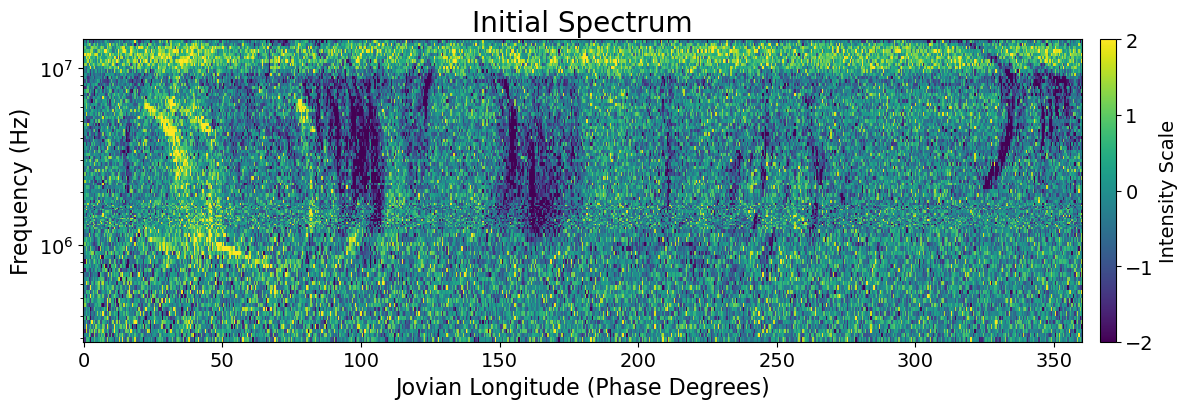}
    \includegraphics[width=0.85\linewidth]{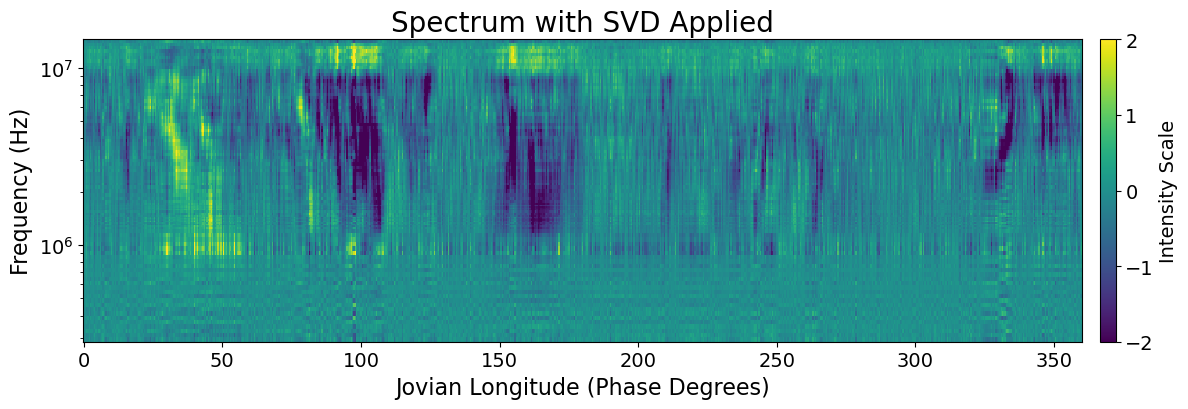}
    \includegraphics[width=0.85\linewidth]{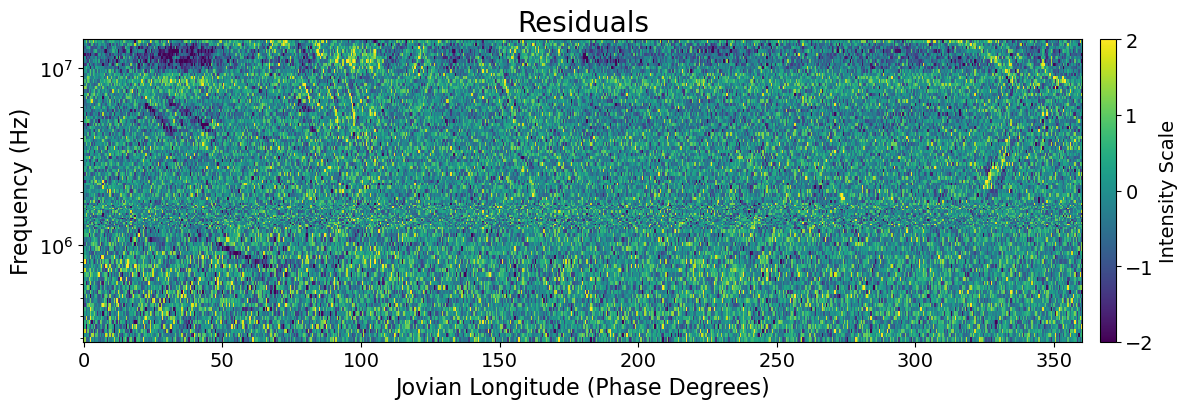}
    \includegraphics[width=0.85\linewidth]{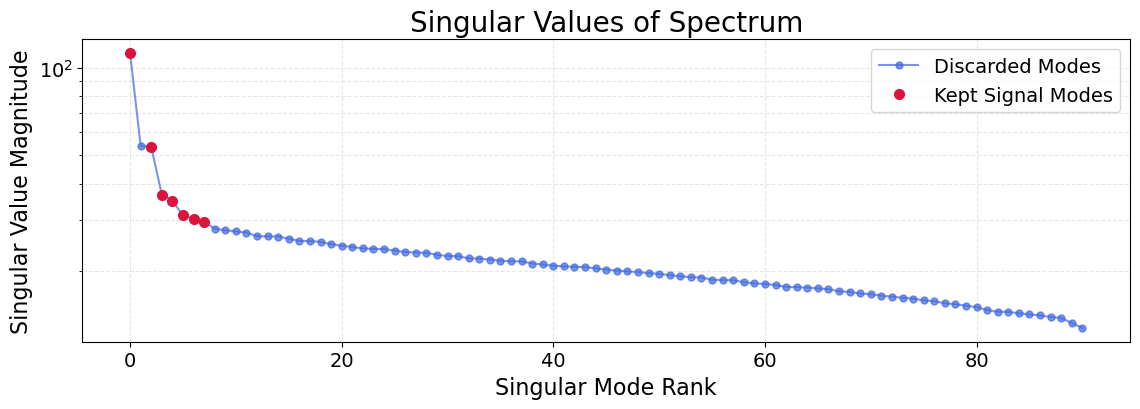}
    \caption{SVD extraction of a series of non-periodic bursts.}
    \label{app3}
\end{figure*}

\begin{figure*}
    \centering
    \includegraphics[width=0.85\linewidth]{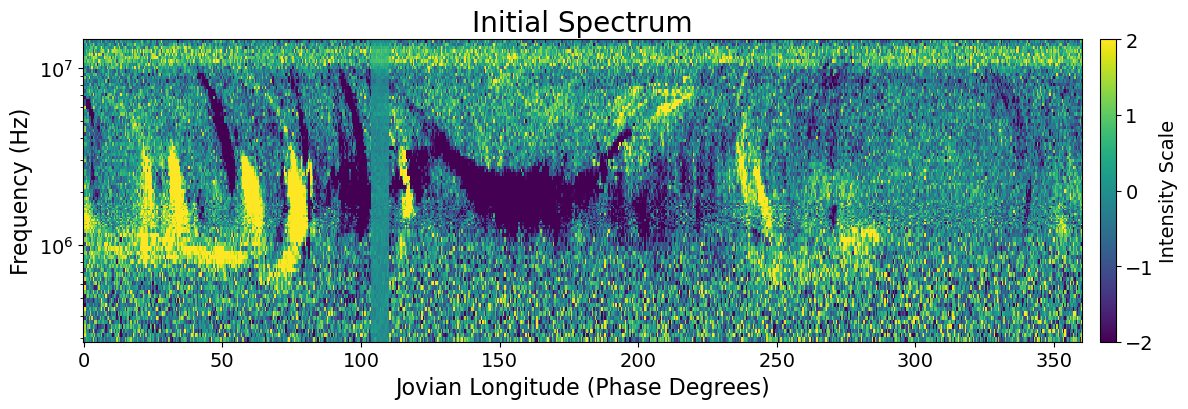}
    \includegraphics[width=0.85\linewidth]{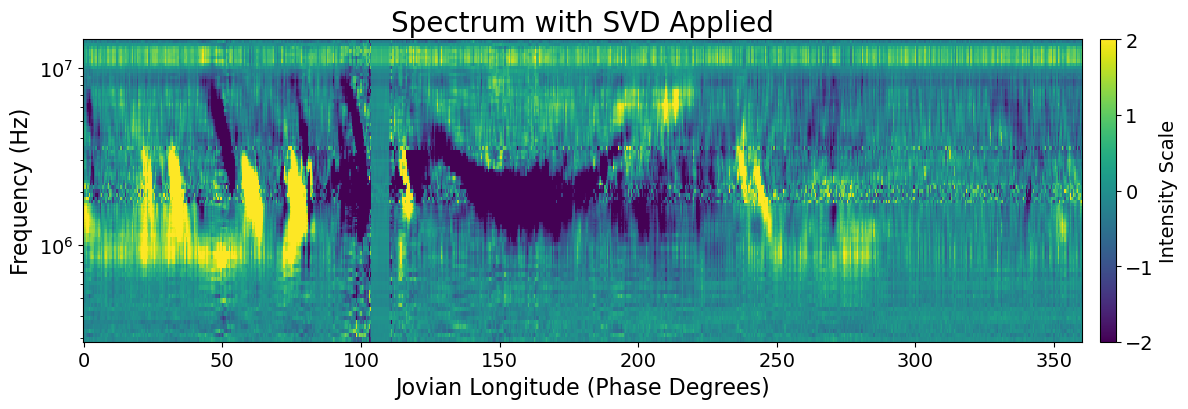}
    \includegraphics[width=0.85\linewidth]{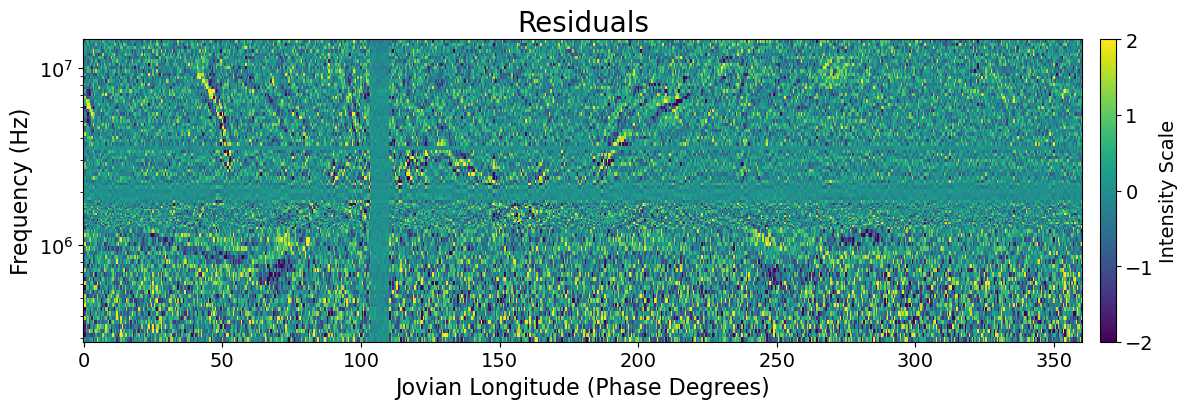}
     \includegraphics[width=0.85\linewidth]{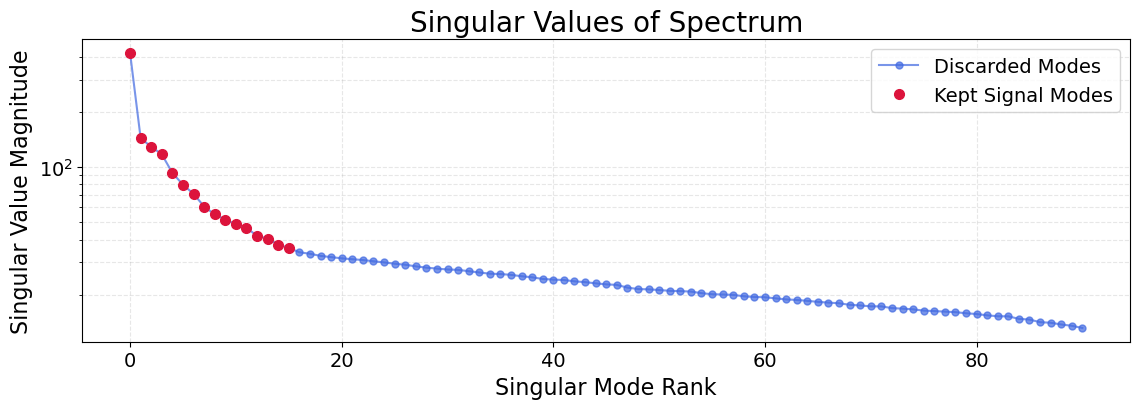}
    \caption{SVD extraction of a strong storm of hectometric arcs.}
    \label{app4}
\end{figure*}

This appendix presents additional representative examples of the SVD extraction pipeline (Section \ref{ssec:svd}) applied across the range of emission morphologies and intensities encountered in the archive, complementing the single event shown in Figure \ref{fig:svd}. Each example follows the same four-panel format: the initial phase-folded spectrum, the spectrum after SVD reconstruction, the residual matrix, and the corresponding singular value distribution with the retained modes indicated. Figure \ref{app1} shows a quieter, two-burst emission; Figure \ref{app2} a vertex-late burst; Figure \ref{app3} a series of non-periodic bursts; and Figure \ref{app4} a strong storm of hectometric arcs.

\end{document}